\documentclass[fleqn,usenatbib]{mnras}

\usepackage{newtxtext,newtxmath}
\usepackage[T1]{fontenc}
\usepackage{ae,aecompl}

\usepackage{graphicx}	% Including figure files
\usepackage{amsmath}	% Advanced maths commands
\usepackage{svg}
\usepackage{gensymb}
\usepackage{xspace}
\usepackage[export]{adjustbox} % for aligning figures left or right
\usepackage{xcolor}
\usepackage{orcidlink}

\newcommand{\pmoffs}[2]{^{+ #1}_{- #2}}

\newcommand{\Msun}{\mbox{$M_{\sun}$}}
\newcommand{\Lsun}{\mbox{$L_{\sun}$}}

\newcommand{\Mjup}{\mbox{$M_{\rm Jup}$}}

\newcommand{\kms}{\mbox{km\,s$^{-1}$}}
\newcommand{\masyr}{\hbox{mas\,yr$^{-1}$}}

\newcommand{\Lbol}{\mbox{$L_{\rm bol}$}}
\newcommand{\Teff}{\mbox{$T_{\rm eff}$}}

\newcommand{\Sinit}{\mbox{$S_{\rm init}$}}
\newcommand{\kbb}{\mbox{$k_{\rm B}$\,baryon$^{-1}$}}
\newcommand{\mur}{\mbox{${\mu_{\alpha^*}}$}}
\newcommand{\mud}{\mbox{${\mu_{\delta}}$}}
\newcommand{\tref}{\mbox{$t_{\rm ref}$}}

\newcommand{\Hipparcos}{{\sl Hipparcos}\xspace}
\newcommand{\Gaia}{{\sl Gaia}\xspace}
\newcommand{\JWST}{{\sl JWST}}

\title[Mass and Initial Entropy of 51 Eri b]{Limits on the Mass and Initial Entropy of 51~Eri~b from Gaia~EDR3 Astrometry}

\author[Dupuy, Brandt, \& Brandt]{
Trent J.~Dupuy,$^{1}$\thanks{E-mail: tdupuy@roe.ac.uk}\orcidlink{0000-0001-9823-1445}
G.~Mirek Brandt,$^{2}$\thanks{NSF Graduate Research Fellow}\orcidlink{0000-0003-0168-3010}
and Timothy D.~Brandt$^{3}$\orcidlink{0000-0003-2630-8073}
\\
$^{1}$Institute for Astronomy, University of Edinburgh, Royal Observatory, Blackford Hill, Edinburgh, EH9 3HJ, UK\\
$^{2}$Department of Physics, University of California, Santa Barbara, Santa Barbara, CA 93106, USA
}

\pubyear{2021}

\begin{document}
\label{firstpage}
\pagerange{\pageref{firstpage}--\pageref{lastpage}}
\maketitle

\begin{abstract}
51~Eri~b is one of the only planets consistent with a wide range of possible initial entropy states, including the cold-start scenario associated with some models of planet formation by core accretion. The most direct way to constrain the initial entropy of a planet is by measuring its luminosity and mass at a sufficiently young age that the initial conditions still matter. We present the tightest upper limit on 51~Eri~b's mass yet ($M<11$\,\Mjup\ at 2$\sigma$) using a cross-calibration of \Hipparcos\ and \Gaia~EDR3 astrometry and the orbit-fitting code {\tt orvara}. We also reassess its luminosity using a direct, photometric approach, finding $\log(\Lbol/\Lsun) = -5.5\pm0.2$\,dex. Combining this luminosity with the $24\pm3$\,Myr age of the $\beta$~Pic moving group, of which 51~Eri is a member, we derive mass distributions from a grid of evolutionary models that spans a wide range of initial entropies. We find that 51~Eri~b is inconsistent with the coldest-start scenarios, requiring an initial entropy of $>8$\,\kbb\ at 97\% confidence. This result represents the first observational constraint on the initial entropy of a potentially cold-start planet, and it continues the trend of dynamical masses for directly imaged planets pointing to warm- or hot-start formation scenarios.
\end{abstract}


%%%%%%%%%%%%%%%%% BODY OF PAPER %%%%%%%%%%%%%%%%%%

\section{Introduction} \label{sec:intro}

Until recently, directly imaged planets have had one key disadvantage compared to their counterparts detected via other methods. Their masses were not directly measurable, unlike planets detected via radial velocities (RVs) or transits. Measuring the mass of a directly imaged planet requires precise astrometric or radial-velocity monitoring of the host star over a large fraction of the planet's orbit. Very long orbital periods and small planet-to-star mass ratios have made this impracticable. 

Absolute astrometry from \Hipparcos and \Gaia have completely changed this picture. We can now detect long-term astrometric signals with a precision of $\sim$100\,$\mu$as or better, corresponding to host-star accelerations of $\lesssim$10\,$\mu$as\,yr$^{-2}$. By the direct measurement of the gravitational pull of the planet on the host star, the mass of the orbiting body can be measured without precise determination of its orbital parameters. This has opened the door for direct mass measurements of brown dwarf companions like Gl~229~B that have orbital periods of hundreds of years \citep{2020AJ....160..196B,Brandt2021_Six_Masses}, as well as short-period planets like $\beta$~Pic~b  \citep{2018NatAs.tmp..114S,Dupuy+Brandt+Kratter+etal_2019}, $\beta$~Pic~c \citep{Brandt_2021_beta_Pic_bc}, and HR~8799~e \citep{Brandt_2021_HR8799e}.

For gas-giant planets young enough to have retained properties related to their formation, mass measurements can potentially provide information about the formation process. In a direct collapse scenario commonly associated with forming via gravitational instability in a gas disk \citep[e.g.,][]{1998ApJ...503..923B}, the planet will inherit most of the specific entropy from its natal gas, resulting in the hottest possible planet at a given mass (``hot start''). All planets with directly measured masses and luminosities to date have been consistent with hot-start models, even those as young as the T~Tauri phase \citep{2019ApJ...878L..37F}.

On the other hand, some core-accretion scenarios \citep[e.g.,][]{Hubickyj_etal_2005} can involve significant loss of entropy during gas accretion \citep{2007ApJ...655..541M}, resulting in a much fainter planet for a given amount of accreted gas (``cold start''). More modern core accretion models strongly disfavour the coldest scenarios \citep[e.g.,][]{2017ApJ...834..149B}, but there is very little direct observational evidence available.

51~Eri~b is one of the coldest planets detected via direct imaging \citep{2015Sci...350...64M}. As such, it is a rare case of a young planet where both hot- and cold-start formation scenarios are consistent with its present-day luminosity, with the next best example being COCONUTS-2~b \citep{2021ApJ...916L..11Z}. Indeed, 51~Eri~b holds the distinction of having the most disparate luminosity-based mass estimates from these two scenarios. As a member of the $\beta$~Pic moving group (BPMG) its age is known well \citep[$24\pm3$\,Myr;][]{2015MNRAS.454..593B}. If 51~Eri~b retained most of the entropy of its birth material, then its present-day low temperature would require that it formed from no more than $\approx$2\,\Mjup\ of gas. If instead its formation allowed it to shed entropy through accretion shocks, then it could harbour as much as $\approx$12\,\Mjup\ of gas.

Here we present an upper limit of the mass of 51~Eri~b based on a new analysis incorporating \Gaia~EDR3 astrometry. In contrast to prior work, we find a good agreement between relative astrometry from direct imaging and the absolute astrometry from \Hipparcos\ and \Gaia. We demonstrate that 51~Eri~b is highly unlikely to be harbouring the amount of gas that has been suggested under the cold-start scenario. We use our measured mass distribution to constrain the initial entropy of 51~Eri~b and, finally, discuss the implications for its formation history.

\section{Astrometric analysis} \label{sec:astrom}

In our astrometric analysis, we used the three proper motions reported for 51~Eri (HIP~21547) in the \Hipparcos-\Gaia\ Catalog of Accelerations \citep[HGCA;][]{Brandt_2018,Brandt_2021_HGCA_EDR3}. The HGCA is a cross-calibration that places the two catalogues in a common inertial frame, and we used the most recent version based on \Gaia~EDR3 \citep{Brandt_2021_HGCA_EDR3}. Briefly, the cross-calibration includes both global (all-sky) and local linear transformations to bring positions and proper motions into agreement, as well as error inflation for both \Hipparcos\ and \Gaia\ astrometry to ensure that proper motion differences across the sky follow appropriate normal distributions centred at zero. The HGCA uses a 60/40-weighted combination of the \Hipparcos\ re-reduction by \citet{2007A&A...474..653V} and the original catalogue by \citet{1997ESASP1200.....E}, respectively, because it matches the long-term proper motions between \Hipparcos and \Gaia better.

\begin{table}
\centering
\caption[]{HGCA-EDR3 Astrometry of 51~Eri (HIP~21547).} \label{tbl:hgca}
\begin{tabular}{lcc}
\hline
Parameter & \Hipparcos & \Gaia\ EDR3 \\
\hline
R.A.\ epoch (yr)  & 1990.9139        & 2015.7606 \\[2pt]
Dec.\ epoch (yr)  & 1990.7250        & 2015.8610 \\[2pt]
$\varpi$ (mas)    &  $33.83\pm0.57$              & $33.439\pm0.078$             \\[2pt]
$\mur$ (\masyr)   &  $43.67\pm0.61$              &  $44.05\pm0.14$              \\[2pt]
$\mud$ (\masyr)   & $-64.20\pm0.48\hphantom{-}$  & $-64.03\pm0.10\hphantom{-}$  \\[2pt]
Corr$(\mur,\mud)$ & $0.0142$                     & $-0.4534\hphantom{-}$        \\[2pt]
H-G $\mur$ (\masyr)   & \multicolumn{2}{c}{ $44.257\pm0.019$                  } \\[2pt]
H-G $\mud$ (\masyr)   & \multicolumn{2}{c}{$-64.177\pm0.011\hphantom{-}$      } \\[2pt]
H-G Corr$(\mur,\mud)$ & \multicolumn{2}{c}{$-0.1266\hphantom{-}$              } \\[2pt]
HGCA-EDR3 $\chi^2$    & \multicolumn{2}{c}{2.997                              } \\
\hline
\end{tabular}
\begin{list}{}{}
\item[Note.] The central epochs of R.A.\ and Dec.\ measurements (see Equation~(1) of \citealp{Brandt_2021_HGCA_EDR3}) are quoted here for plotting purposes. Our orbital analysis models the individual epoch astrometry using {\tt htof} \citep{Brandt_2021_htof}.
\end{list}
\end{table}

\begin{table}
\setlength{\tabcolsep}{3pt}
\centering
\caption[]{Orbital parameters derived from the joint fitting of HGCA-EDR3 proper motions and relative astrometry.} \label{tbl:orbit}
\begin{tabular}{lcc}
\hline
Parameter & Median $\pm$ 1$\sigma$ & 95.4\% c.i. \\
\hline
Mass of the planet $M_{\rm pl}$ (\Mjup)                  & $3.8_{-3.8}^{+1.8**}$      &      0.0, 10.9      \\[3pt]
Mass of the host star $M_{\star}$ (\Msun)                & $1.75\pm0.05^{**}$         &     1.65, 1.85      \\[3pt]
Semimajor axis $a$ (AU)                                  & $10.4_{-1.1}^{+0.8}$       &      8.8, 14.9      \\[3pt]
Eccentricity $e$                                         & $0.57_{-0.06}^{+0.08}$     &     0.35, 0.66      \\[3pt]
Inclination $i$ (\degree)                                & $141_{-7}^{+9}$            &      124, 154       \\[3pt]
PA of the ascending node $\Omega$ (\degree)              & $60_{-60}^{+40}$           &        0, 130       \\[3pt]
Argument of periastron $\omega$ (\degree)                & $82\pm23$                  &       49, 120       \\[3pt]
Mean longitude $\lambda_{\rm ref}$ at $\tref^*$ (\degree)& $160\pm50$                 &      100, 290       \\[3pt]
Barycentric \mur\ (\masyr)                               & $44.255\pm0.022$           &    44.23, 44.29     \\[3pt]
Barycentric \mud\ (\masyr)                               &$-64.176\pm0.011\hphantom{-}$&$-$64.22, $-$64.16  \\[3pt]
\hline												                        
Period $P$ (yr)                                          & $25_{-4}^{+3}$             &     19.7, 43.3      \\[3pt]
Time of periastron $t_p$ (Jyr)                            & $2027.8_{-2.2}^{+2.5}$     &   2024.4, 2044.5    \\[3pt]
$\tau \equiv (t_p-\tref)/P^*$                            & $0.78\pm0.10$              &     0.40, 0.94      \\
\hline
\end{tabular}
\begin{list}{}{}
\item[Note.] Free parameters in the MCMC are shown in the top section. These were used to compute the parameters at the bottom. All posterior distributions are non-Gaussian, except $M_{\star}$, while $\omega$, $\Omega$, and $\lambda_{\rm ref}$ are also multi-modal.
\item[*] Reference epoch $\tref = 2010.0$ (55197\,MJD).
\item[**] The posterior of $M_{\rm pl}$ peaks near zero. This result should be used only as an upper limit. The posterior of $M_{\star}$ is identical to the prior we assumed.
\end{list}
\end{table}

Table~\ref{tbl:hgca} presents the absolute astrometry for 51~Eri that we used in our analysis. The key measurements are the three proper motions: the \Hipparcos proper motion (measured near epoch 1991), the \Gaia EDR3 proper motion (measured near epoch 2016), and the long-baseline proper motion derived from the RA and Dec measurements from the two missions (denoted ``H-G''). This final value is by far the most precise because proper motion uncertainties scale inversely with the time baseline, and the time elapsed between the two missions is nearly an order of magnitude longer than either one individually. In addition, 51~Eri is so bright ($G=5.1\,{\rm mag}$) that \Gaia~EDR3 cannot achieve the $\approx$20\,$\mu$as\,yr$^{-1}$ precision it reaches at slightly fainter magnitudes \citep{Lindegren+Klioner+Hernandez+etal_2021}. The low $\chi^2$ of 51~Eri reported in HGCA-EDR3 (equivalent to 1.2\,$\sigma$ difference from constant proper motion) indicates that the three proper motions are statistically in good agreement with each other. Therefore, it is only possible to place an upper limit on the mass of 51~Eri~b.

\begin{figure}
 \begin{center}
  \includegraphics[width=0.49\textwidth]{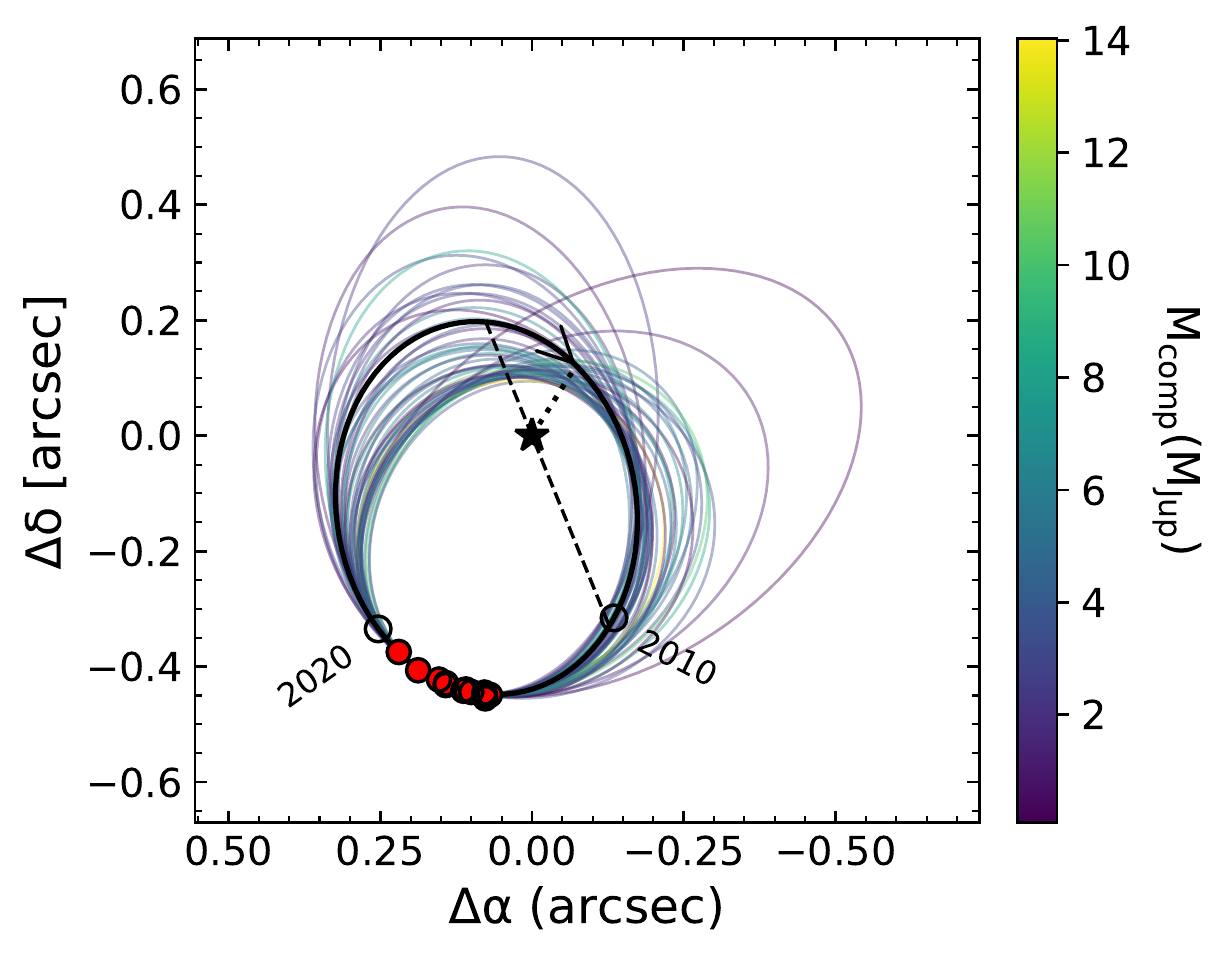}
  \vskip -6pt
  \caption{Relative astrometry (red circles) from \citet{De_Rosa_2020_51Erib} plotted alongside 50 randomly drawn orbits drawn from our MCMC posterior (thin lines colour-coded by planet mass). The maximum-likelihood orbit (thick black line) is also shown with its line of nodes (dashed line), time of periastron passage (dotted line and arrow), and predicted positions at epochs 2010.0 and 2020.0 (open circles). The two modes of the MCMC posterior are evident as a collection of orbits with maximum projected separations either toward the top of the plot or toward the right. These two modes correspond to modes in viewing angles ($\Omega$, $\omega$, and $\lambda_{\rm ref}$) and do not correlate strongly with any physical parameters like $M_{\rm pl}$, $a$, or $e$. \label{fig:skyplot}}
 \end{center}
\end{figure}

\begin{figure*}
 \centerline{\includegraphics[width=0.46\textwidth]{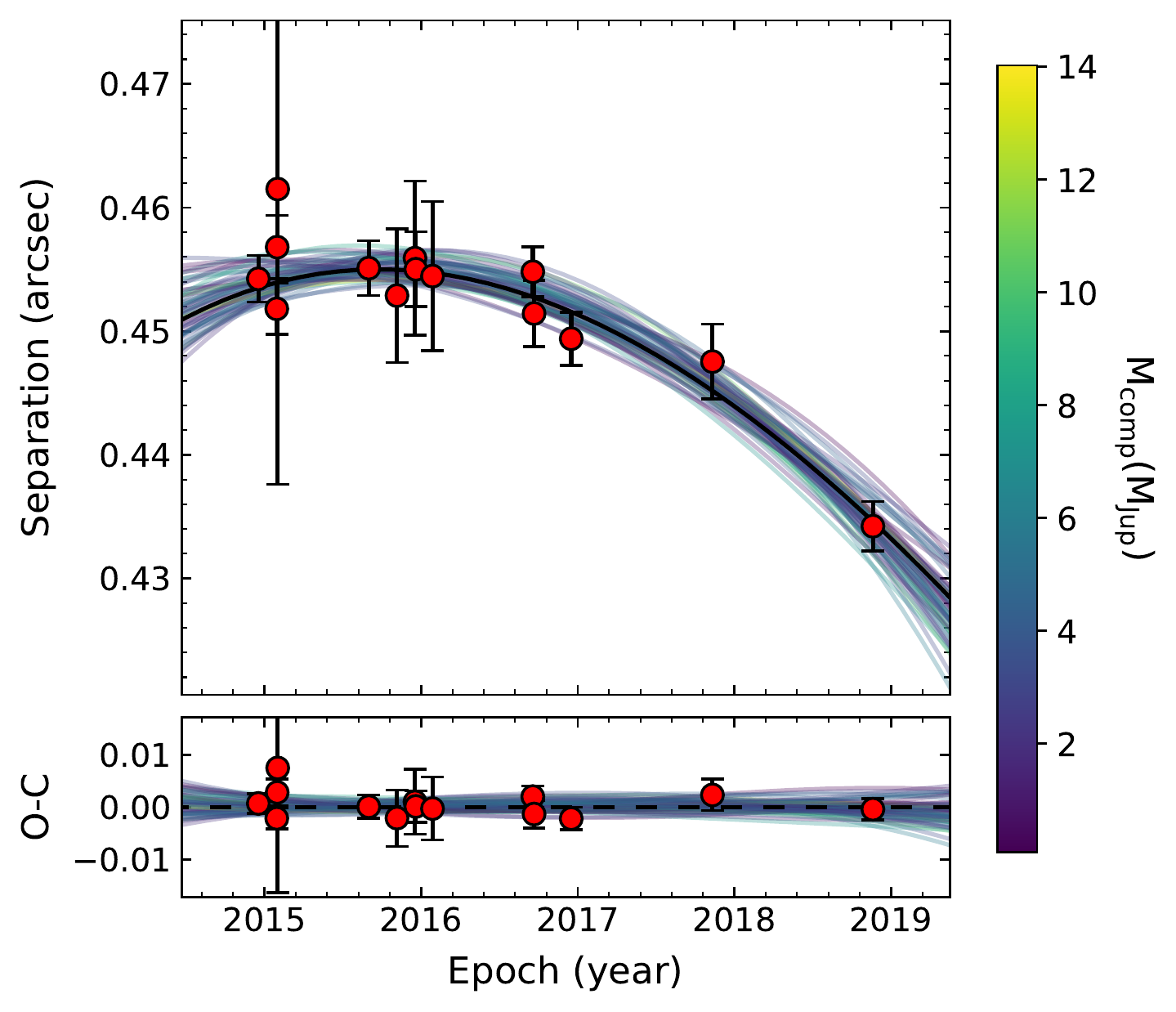} \hfill
             \includegraphics[width=0.46\textwidth]{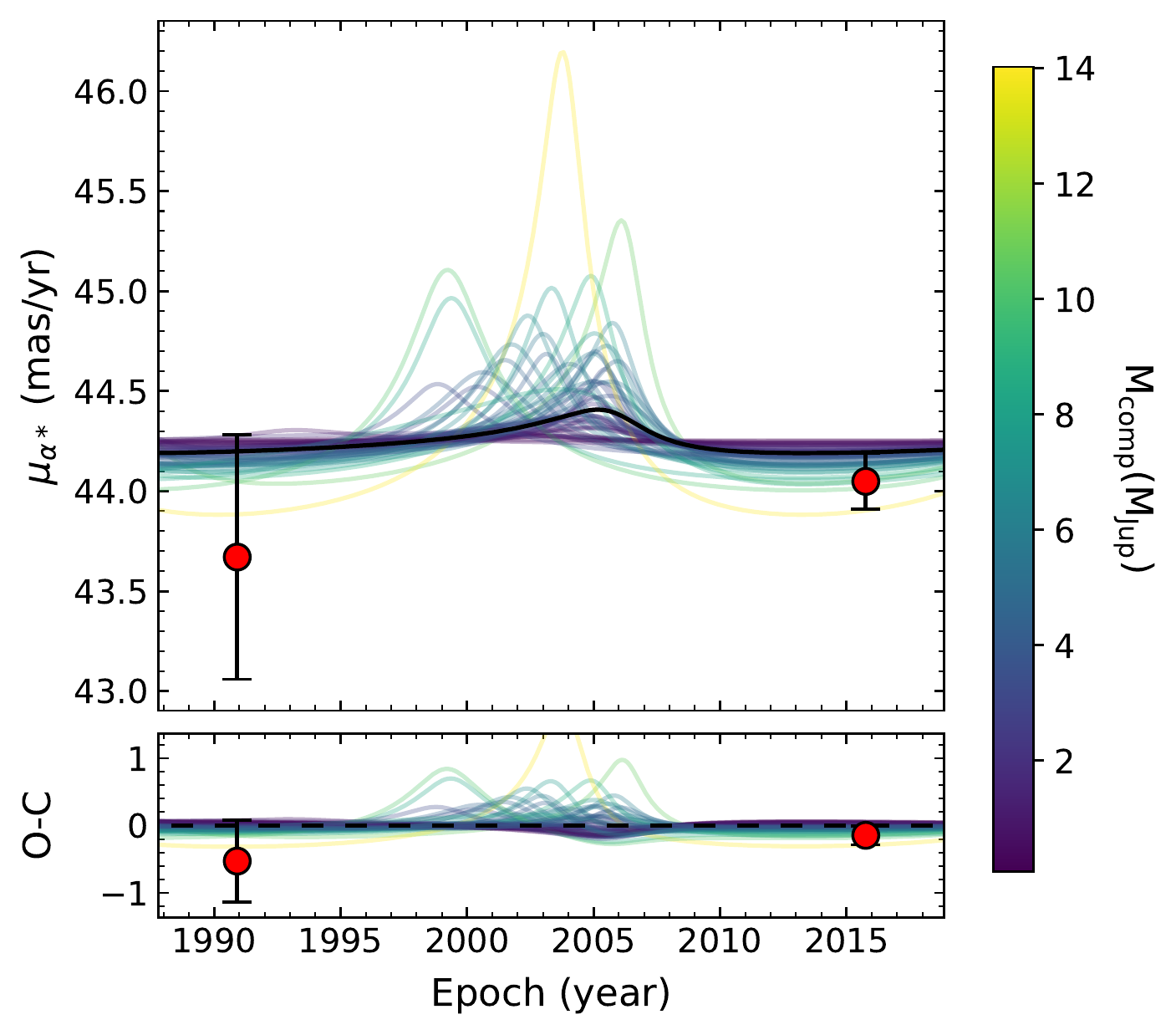}}
 \centerline{\includegraphics[width=0.46\textwidth]{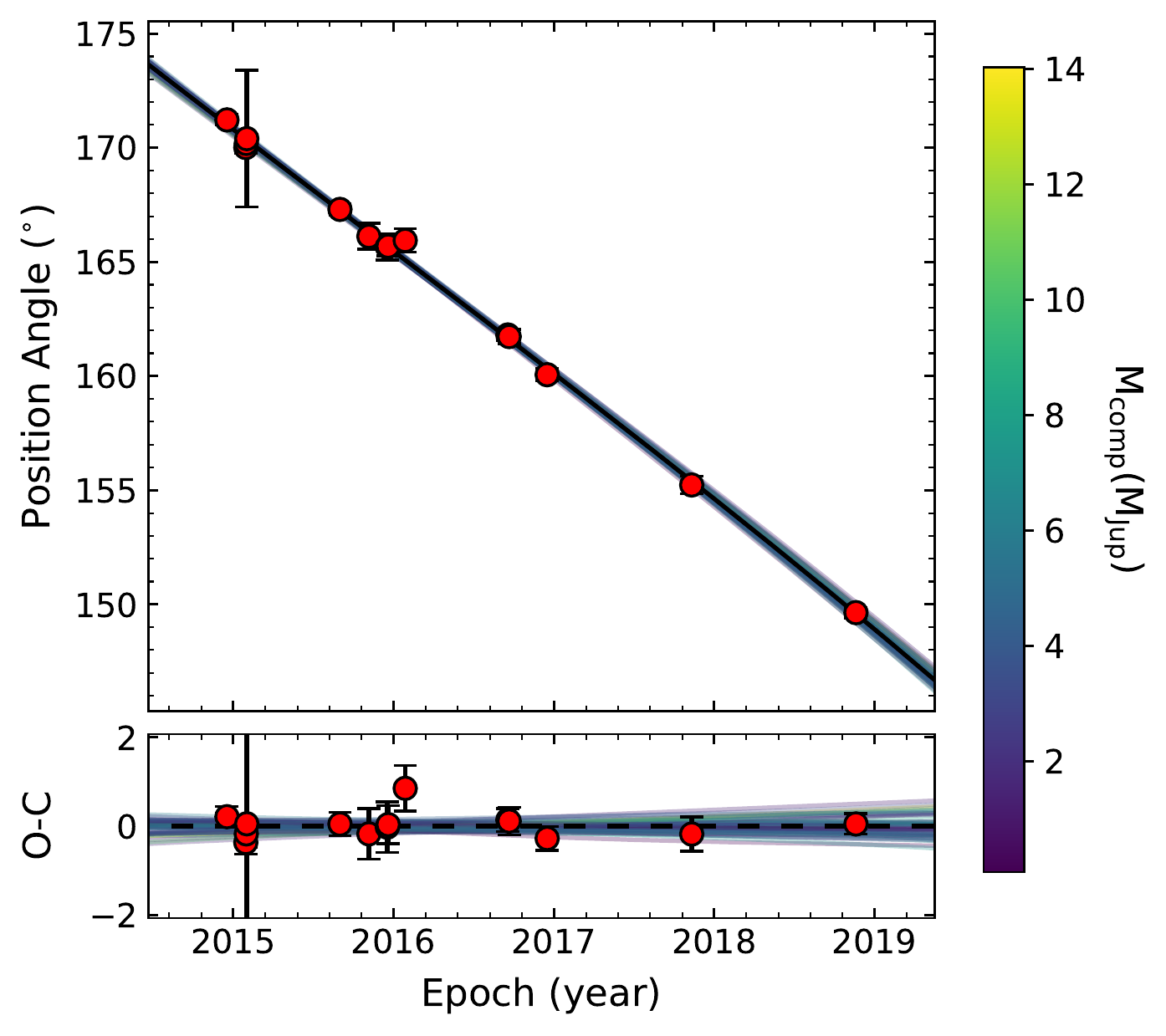} \hfill
             \includegraphics[width=0.46\textwidth]{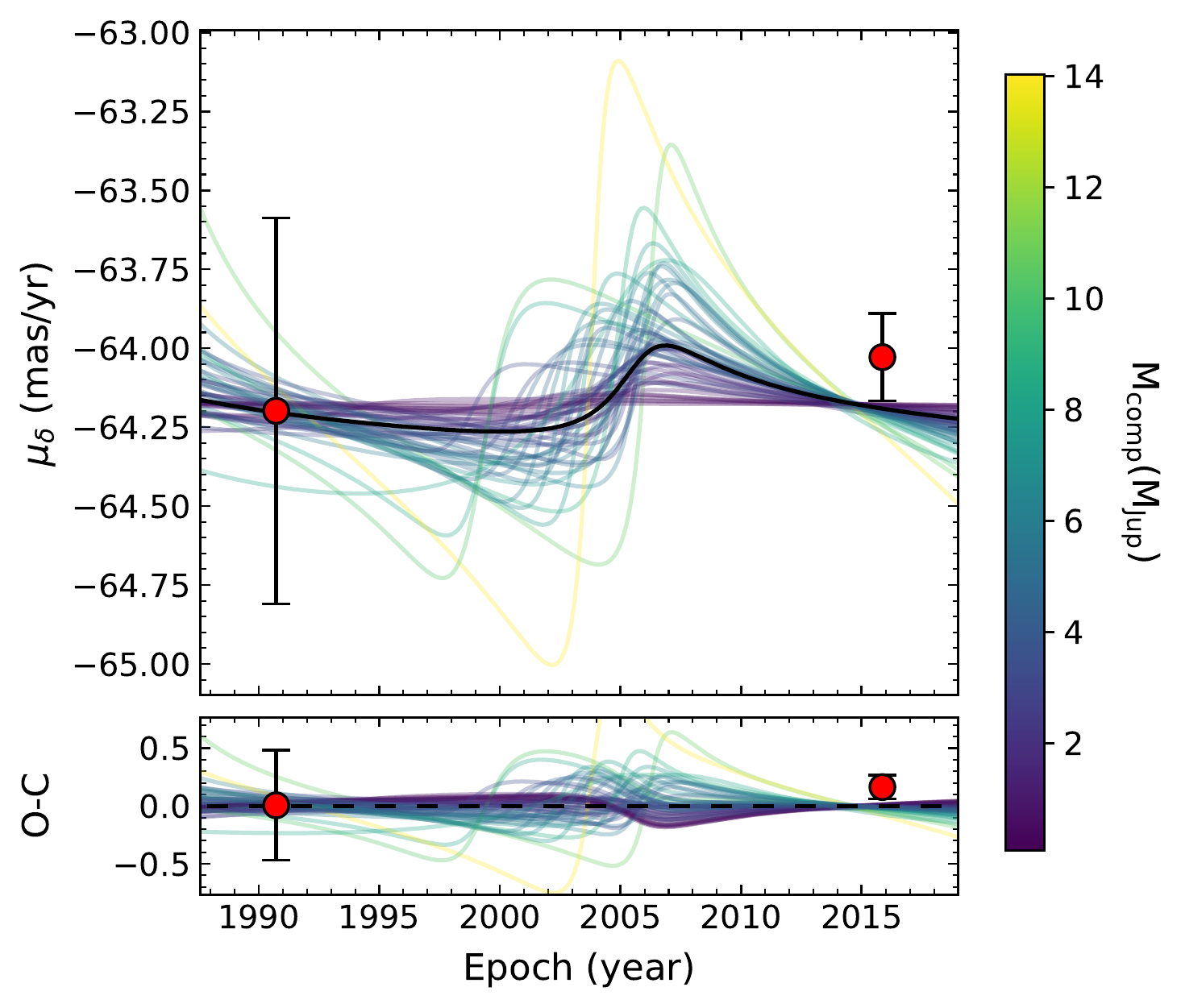}}
  \caption{Relative astrometry (left) from \citet{De_Rosa_2020_51Erib} and proper motions from HGCA-EDR3 (right) shown alongside the same orbits drawn from our MCMC posterior as in Figure~\ref{fig:skyplot} (colour-coded by planet mass). We find that the relative orbit is in excellent agreement with the null detection of a change in proper motion between \Hipparcos\ and \Gaia\ epochs for a wide range of planetary masses. The maximum-likelihood orbit (thick black line) has a period of 25~years, nearly identical to the difference between these two epochs.}
 \label{fig:orbitfit}
\end{figure*}

\begin{figure*}
 \centerline{\includegraphics[width=0.25\textwidth]{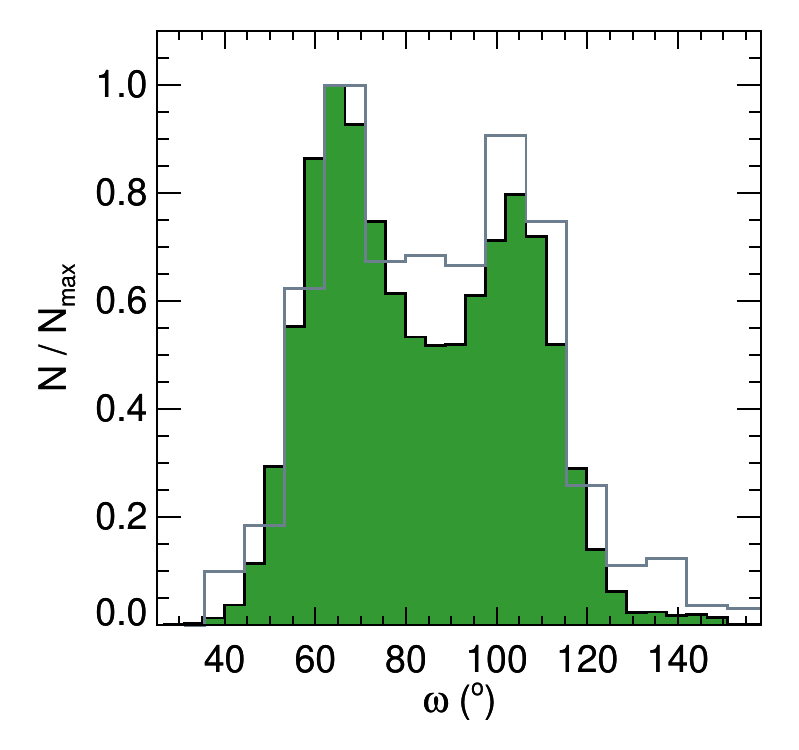}
             \includegraphics[width=0.25\textwidth]{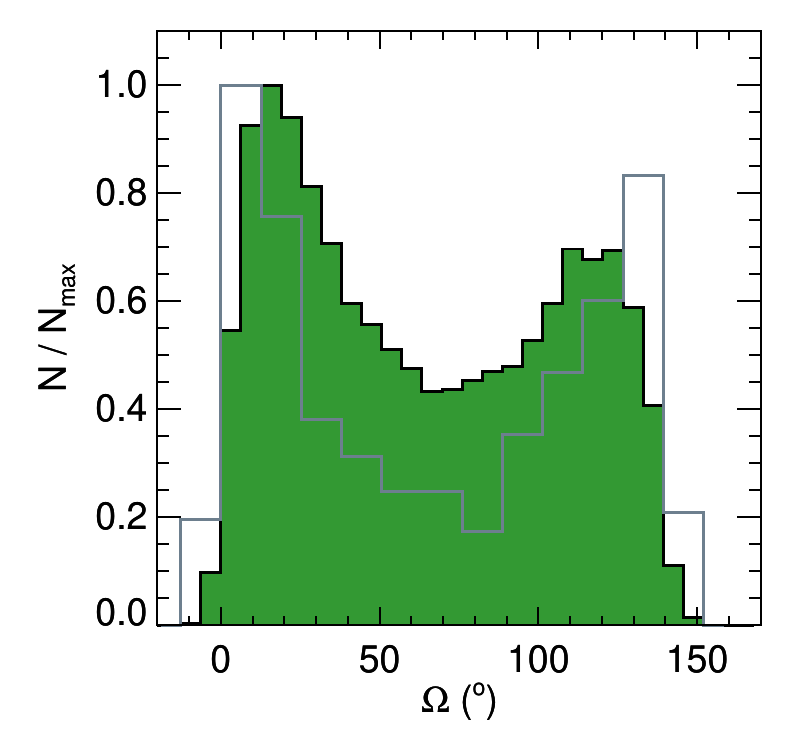}
             \includegraphics[width=0.25\textwidth]{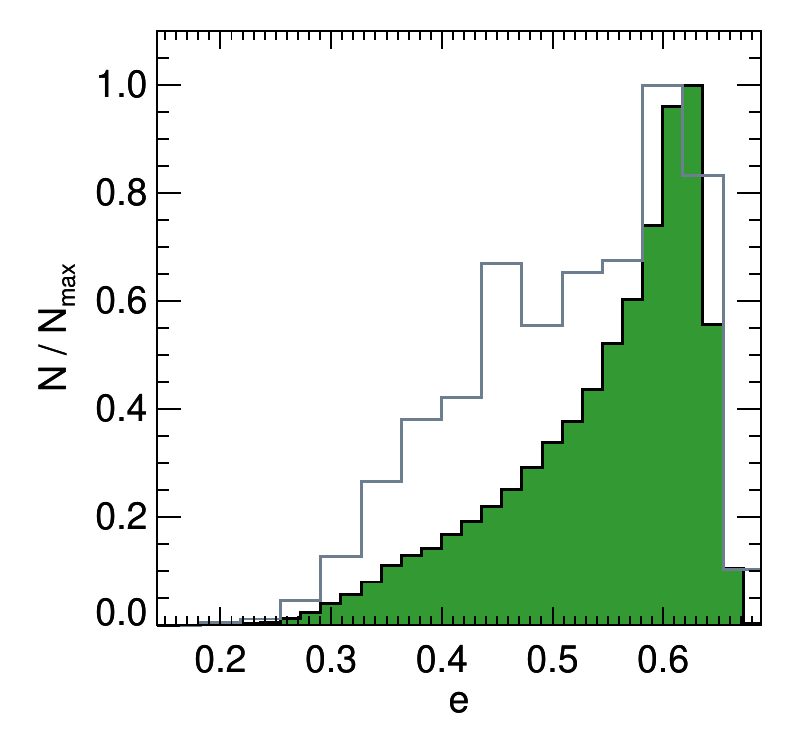}
             \includegraphics[width=0.25\textwidth]{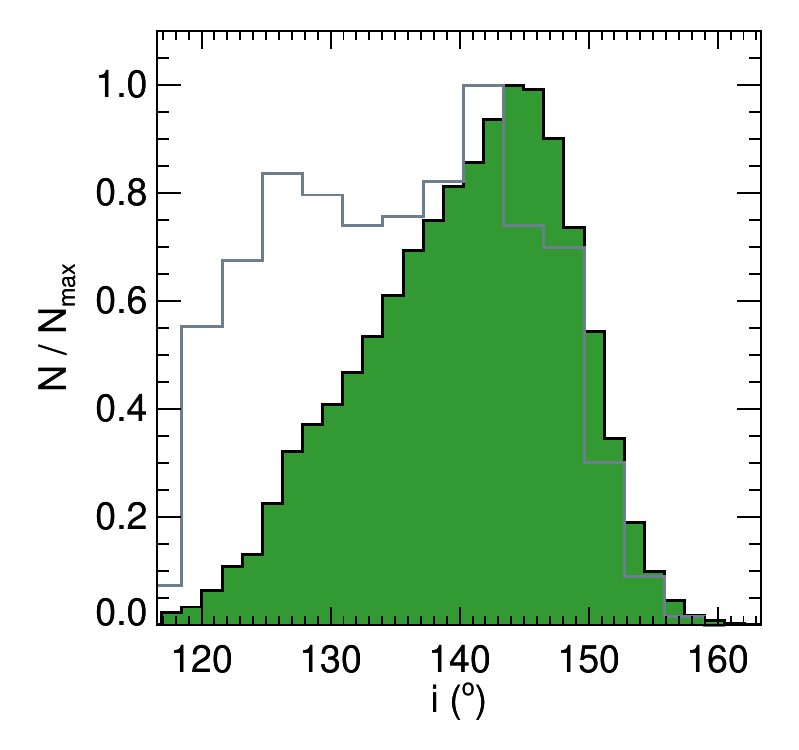}}
 \centerline{\includegraphics[width=0.25\textwidth]{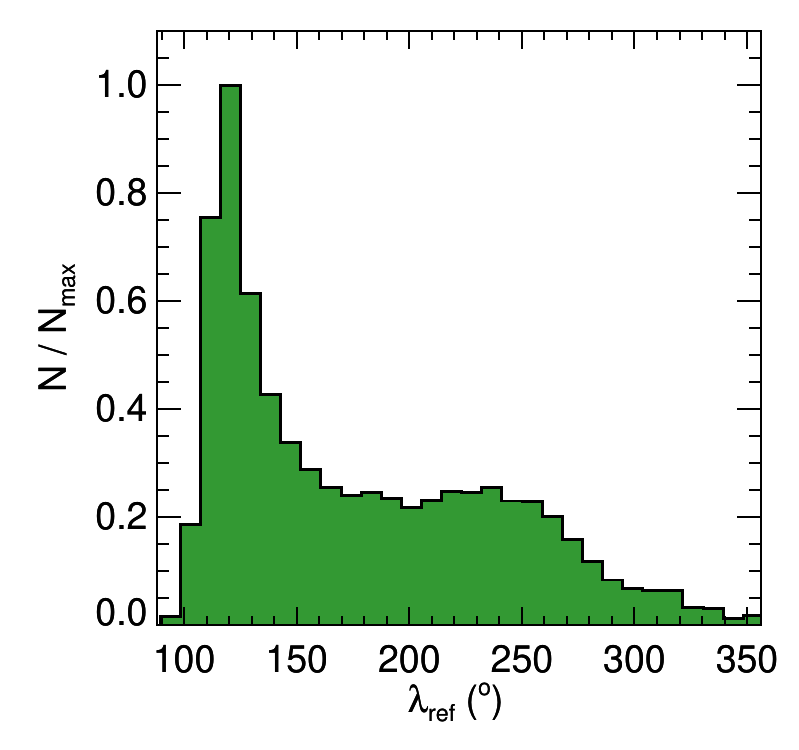}
             \includegraphics[width=0.25\textwidth]{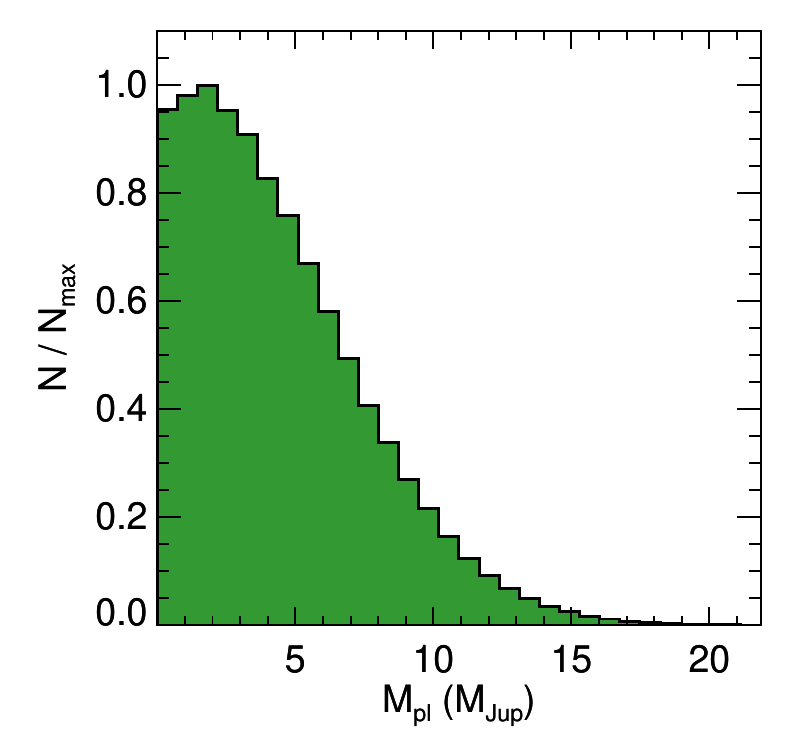}
             \includegraphics[width=0.25\textwidth]{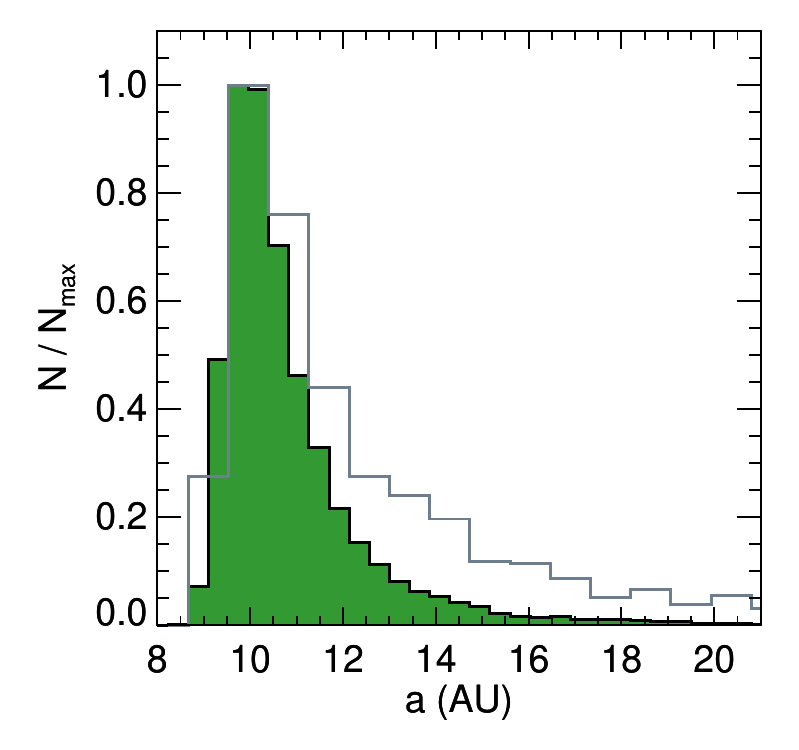}
             \includegraphics[width=0.25\textwidth]{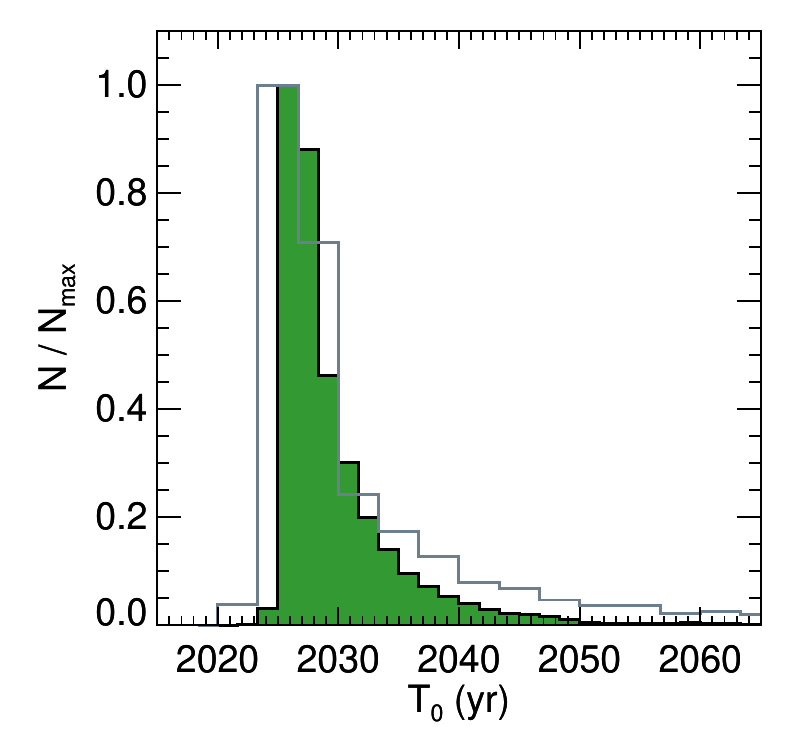}}
  \caption{Histograms of the posterior distributions of orbital parameters fitted in our MCMC analysis (filled, green) shown alongside the posteriors from \citet[][unfilled grey]{De_Rosa_2020_51Erib}. Our results are generally consistent with their analysis based on the same relative astrometry but \Gaia~DR2 proper motions and parallax.}
 \label{fig:orbithist}
\end{figure*}

In order to infer 51~Eri~b's mass, we performed a two-body, joint Keplerian orbit analysis of the HGCA-EDR3 proper motions and the relative astrometry reported by \citet{De_Rosa_2020_51Erib}.
For orbit fitting we used {\tt orvara} \citep[v1.0.4;][]{Brandt_2021_orvara}, which is designed to be used with the HGCA, employs a highly-efficient eccentric anomaly solver, and forward-models epoch astrometry using the {\tt htof} package \citep{Brandt_2021_htof}. 
We inferred posterior distributions via the parallel-tempering fork \citep{Vousden_2016_PT} of the affine-invariant \citep{2010CAMCS...5...65G} Markov-Chain Monte Carlo (MCMC) sampler {\tt emcee} \citep{2013PASP..125..306F}. Our results are based on a run with 100 walkers and $10^6$ steps for the MCMC and 5 temperatures for parallel-tempering. We thinned our chains, retaining every 50th step, and discarded the first 50\% as burn-in, yielding $10^6$ final samples in our posterior.

The eight parameters we fitted were the masses of the host star $M_{\star}$ and planet $M_{\rm pl}$, semimajor axis $a$, inclination $i$, eccentricity $e$, argument of periastron $\omega$, position angle of the ascending node $\Omega$, and mean longitude at the reference epoch of 2010.0 $\lambda_{\rm ref}$. We adopted the normal convention that $0\degree\leq\Omega<180\degree$ in the absence of radial velocity information that is needed to determine whether the ascending node is on the east or west side of the orbit.
For all parameters except the two masses, we adopted the default {\tt orvara} priors: log-flat in $a$, $\sin{i}$ for inclination, and linear-flat in all others. We adopted a Gaussian prior on the primary mass of $1.75\pm0.05$\,\Msun\ and a linear-flat prior on the planet mass ($M_{\rm pl}>0$), rather than log-flat, because we were deriving an upper limit.
{\tt orvara} marginalises over the parallax and barycentric proper motion rather than stepping along these additional dimensions in the MCMC. We adopted a Gaussian prior on the parallax corresponding to the \Gaia~EDR3 measurement (Table~\ref{tbl:hgca}) and an uninformative, uniform prior on the barycentric proper motion.

\begin{figure*}
 \raggedright
 \includegraphics[width=0.190\textwidth]{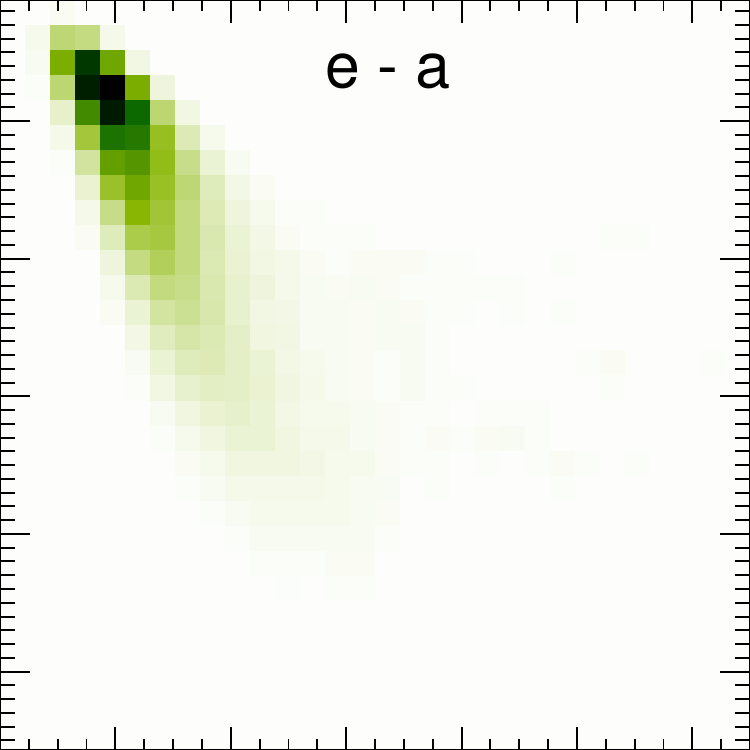} \\
 \includegraphics[width=0.190\textwidth]{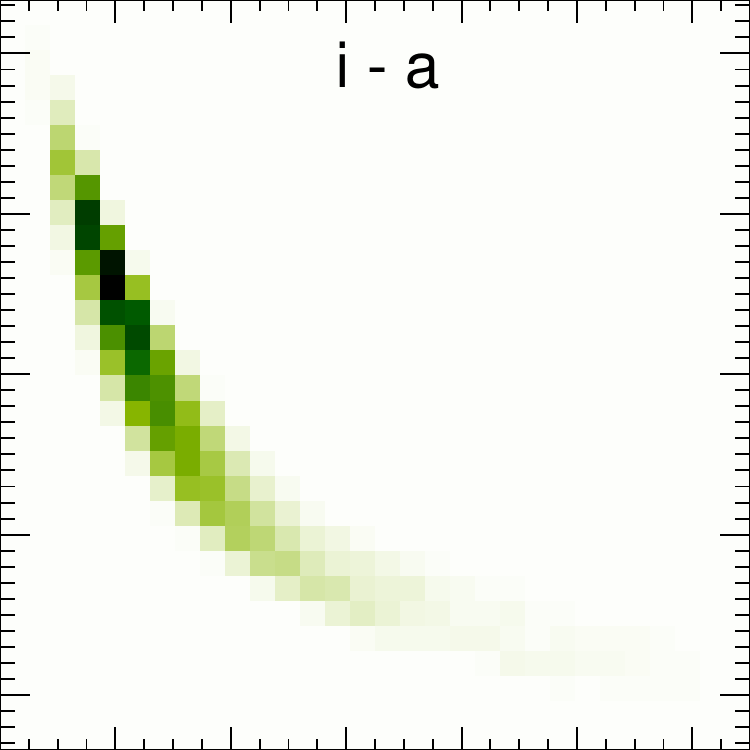}
 \includegraphics[width=0.190\textwidth]{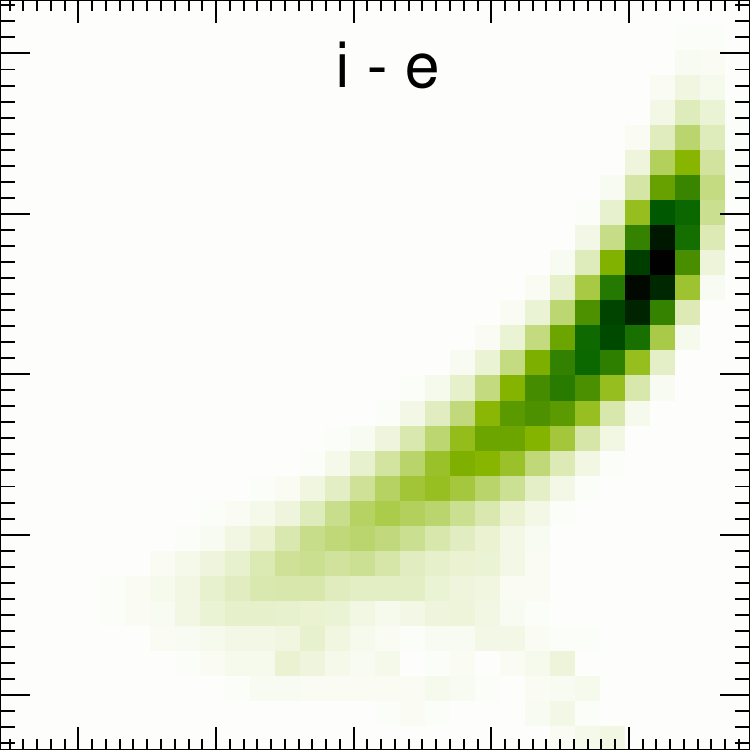} \\
 \includegraphics[width=0.190\textwidth]{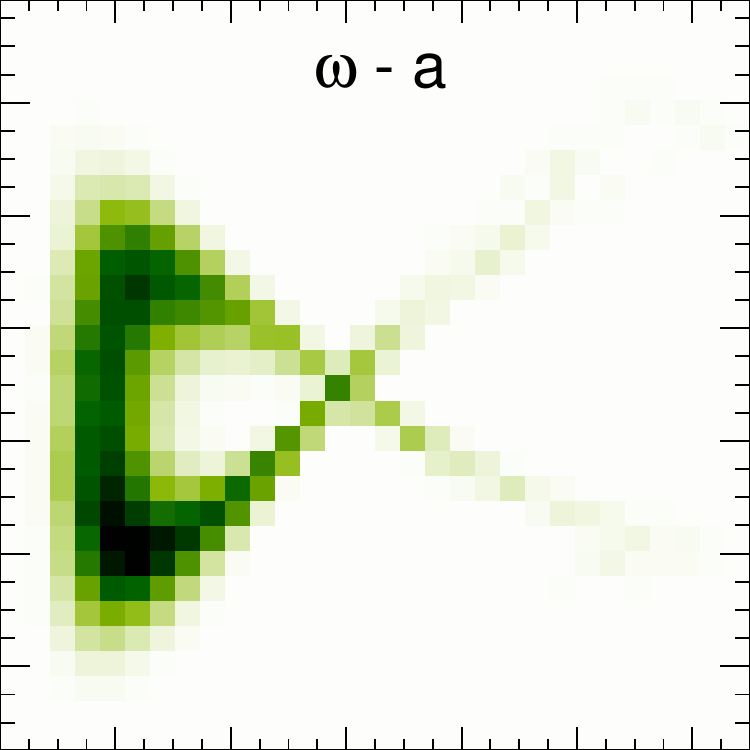}
 \includegraphics[width=0.190\textwidth]{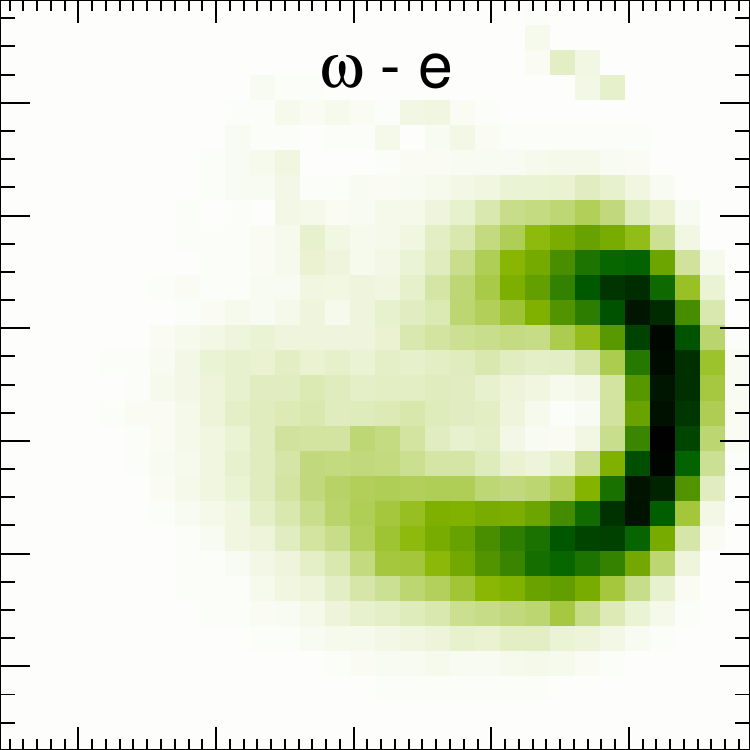}
 \includegraphics[width=0.190\textwidth]{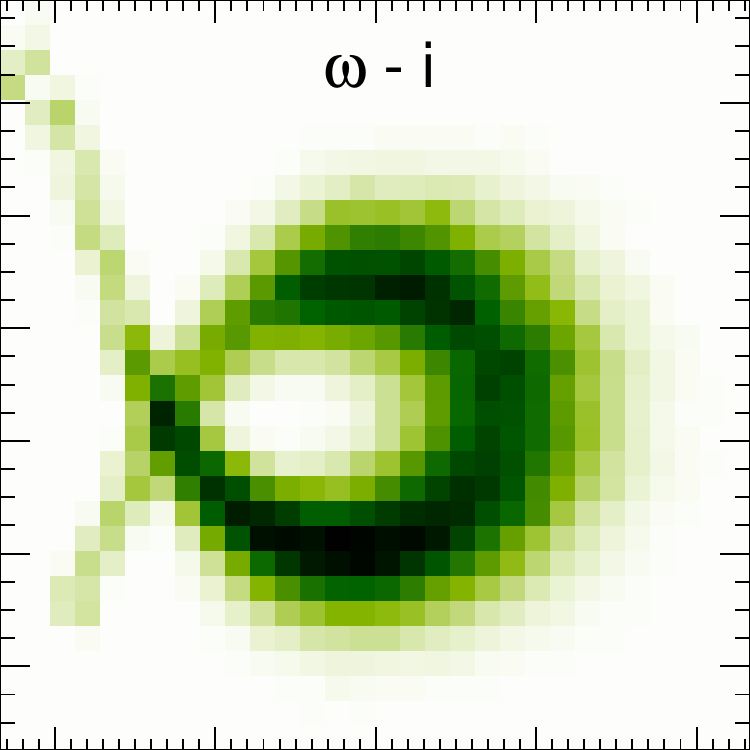} \\
 \includegraphics[width=0.190\textwidth]{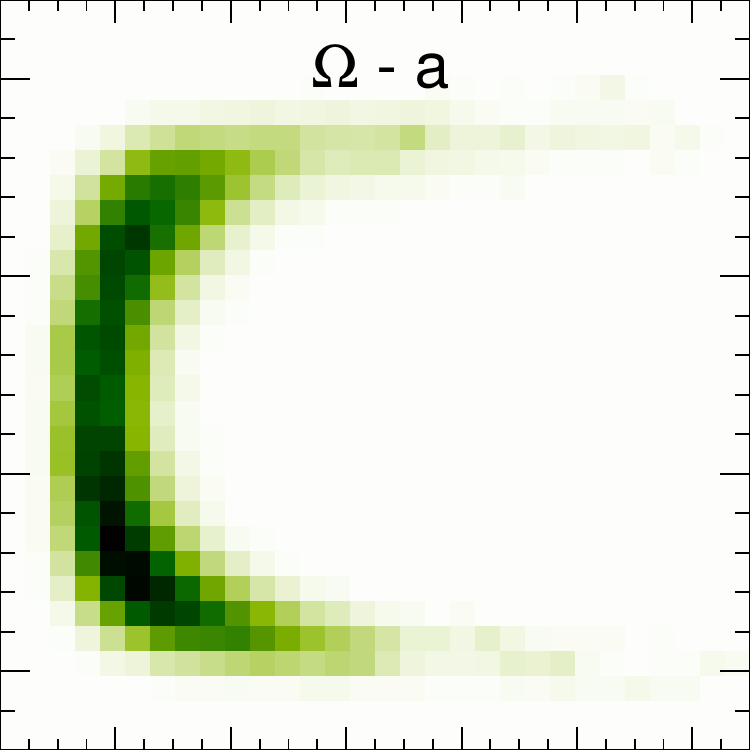}
 \includegraphics[width=0.190\textwidth]{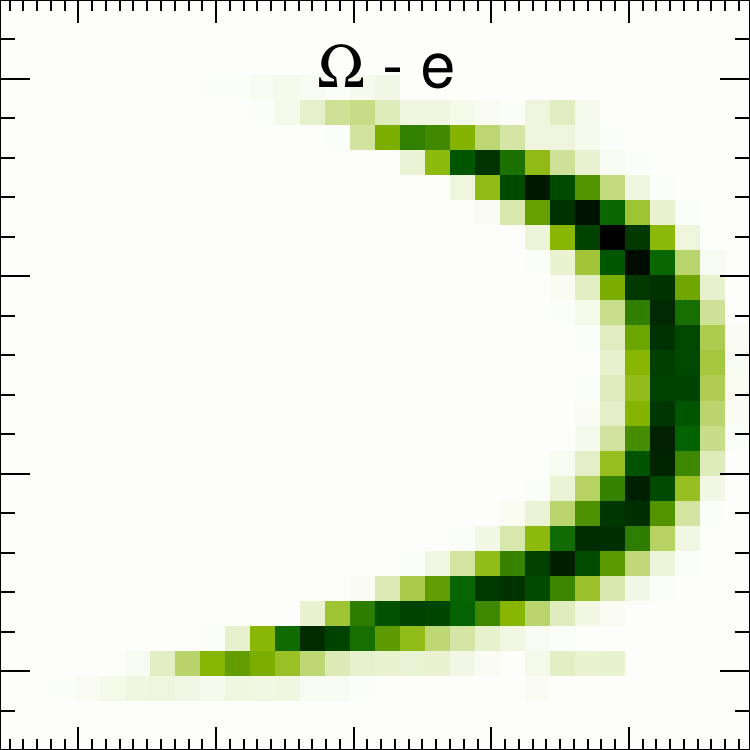}
 \includegraphics[width=0.190\textwidth]{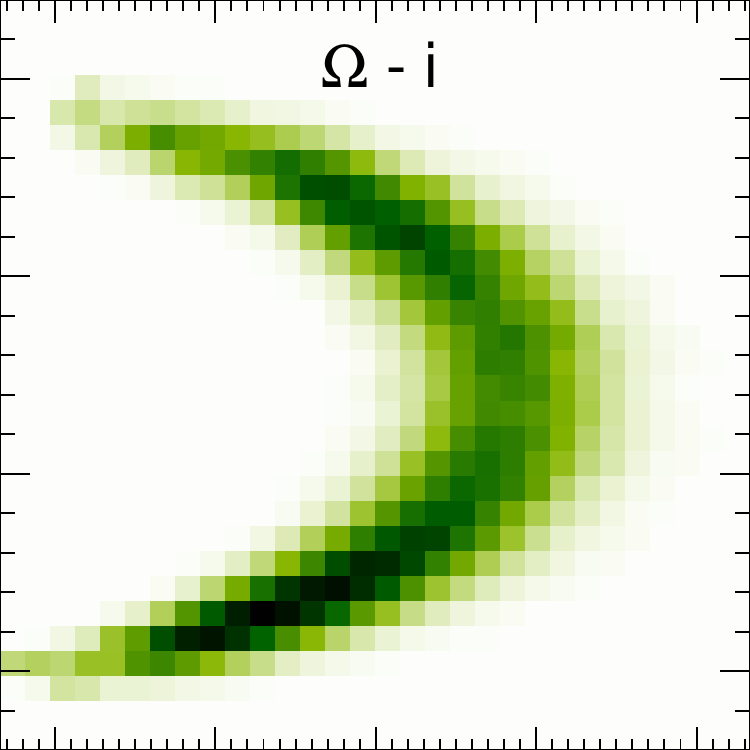}
 \includegraphics[width=0.190\textwidth]{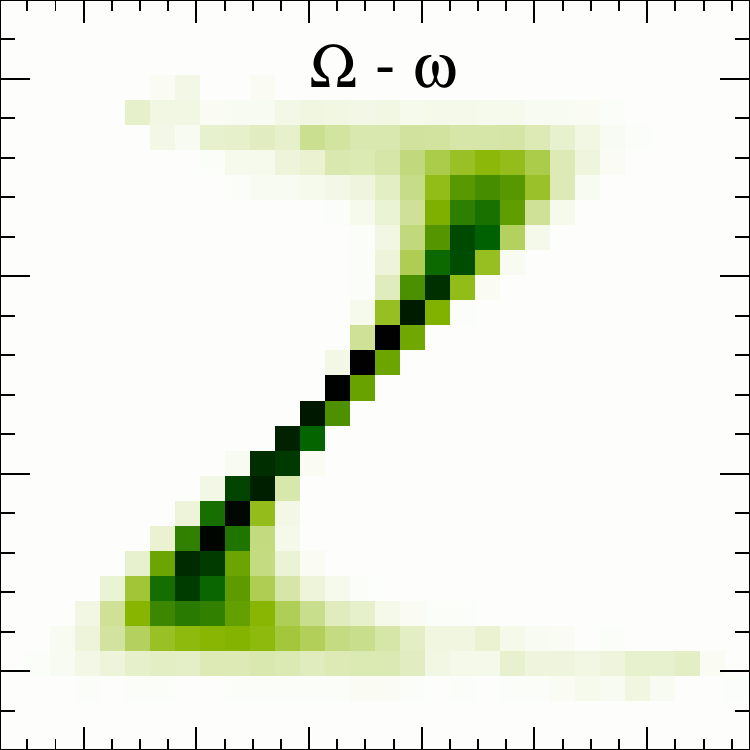} \\
 \includegraphics[width=0.190\textwidth]{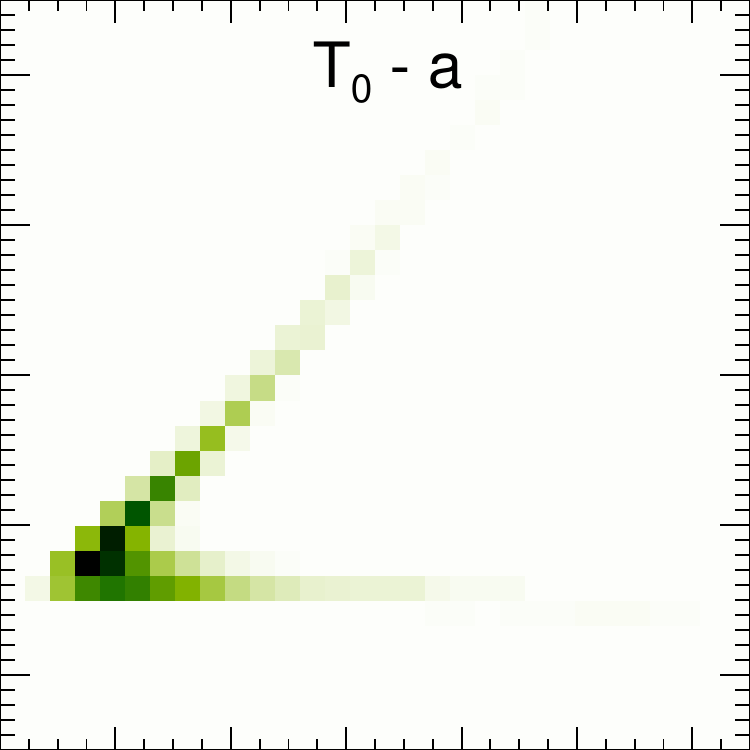}
 \includegraphics[width=0.190\textwidth]{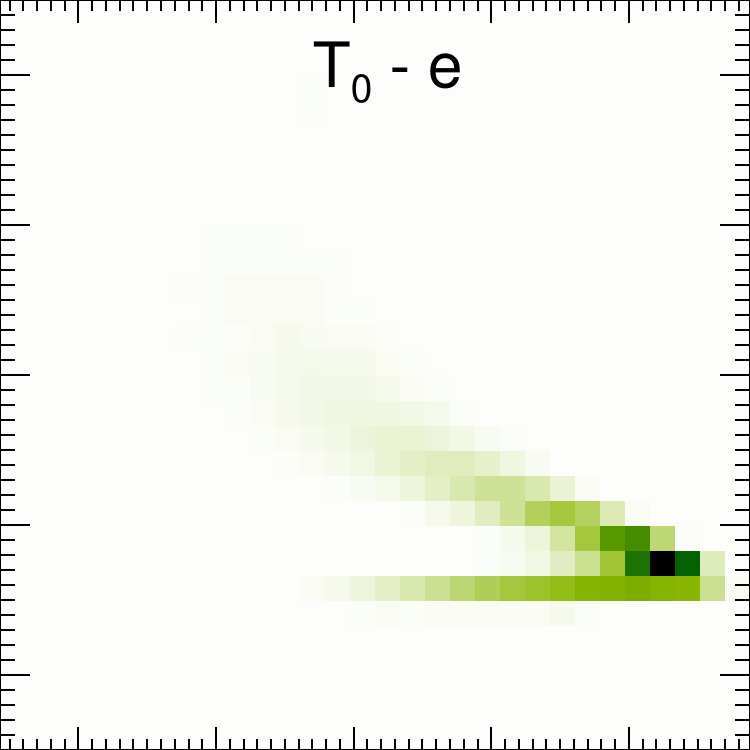}
 \includegraphics[width=0.190\textwidth]{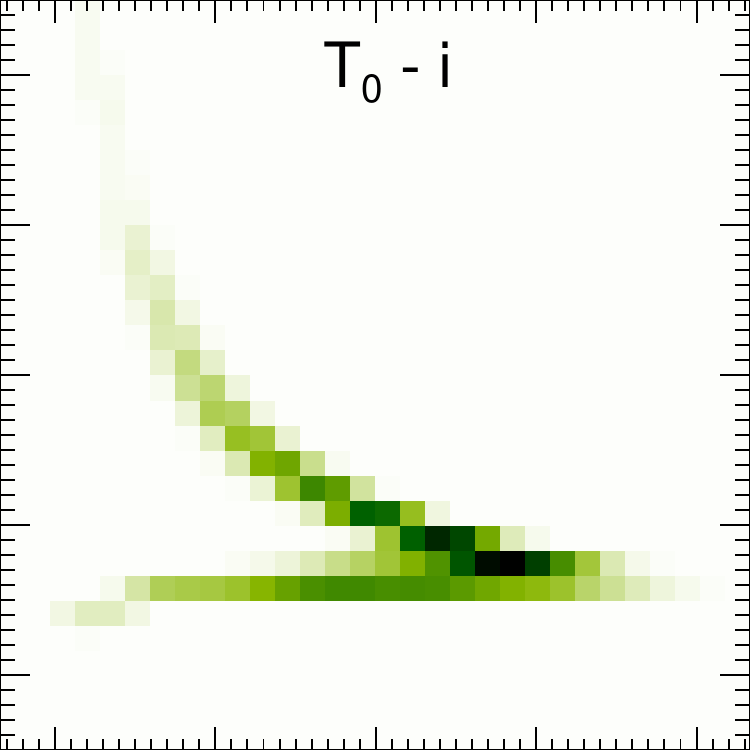}
 \includegraphics[width=0.190\textwidth]{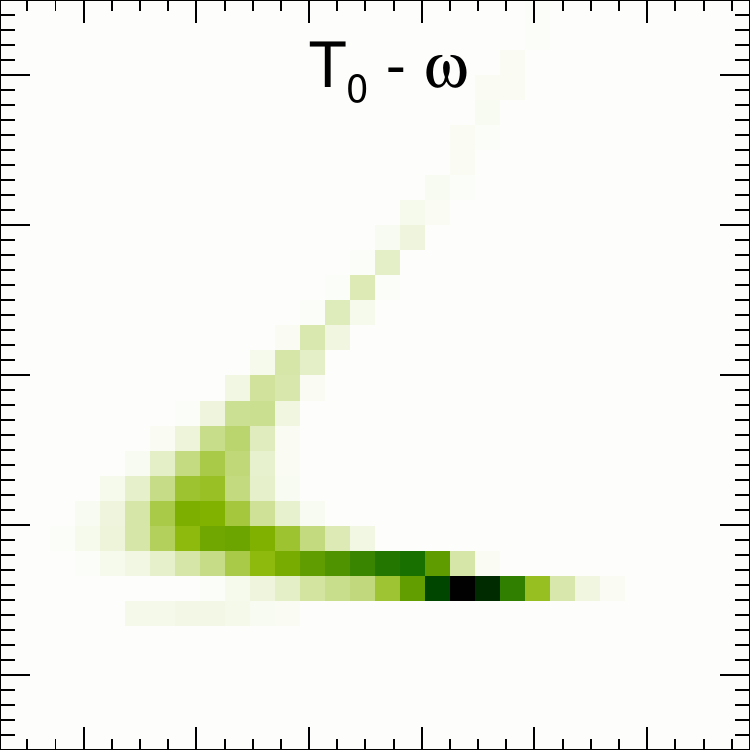}
 \includegraphics[width=0.190\textwidth]{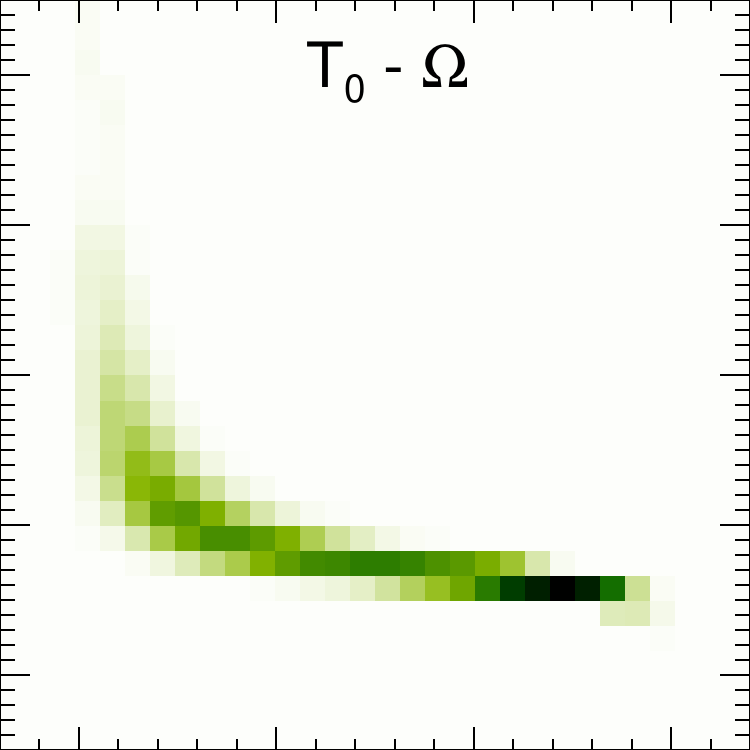}
 \caption{Probability density maps for six independent orbital parameters, excluding the two masses, derived from our MCMC posteriors. Each panel is labelled with the parameters being plotted in the form of $y-x$ axes, and the corresponding axis ranges are identical to the ones used in Figure~\ref{fig:orbithist}. The panel layout matches Figure~3 of \citet{De_Rosa_2020_51Erib} for ready comparison. All of the orbital covariances we find are qualitatively identical to those found in that past work, but with slightly narrower marginalised distributions owing to the improved \Gaia~EDR3 parallax and proper motions now available.}
 \label{fig:corner}
\end{figure*}

Figure~\ref{fig:skyplot} displays the range of astrometric orbits found in our MCMC posterior. We find broadly consistent results as in past work, with two families of orbital solutions that have maximum projected separations toward either the north or to the northwest. These are not the apoastron locations; those are mostly concentrated around PAs of 150--$180\degree$, overlapping almost entirely with the location of 51~Eri~b at the epochs of relative astrometry measurements (2014--2018). Figure~\ref{fig:orbitfit} shows the excellent agreement between the best-fit orbit, the relative astrometry ($\chi^2=13.5$ for 28 measurements), and the HGCA-EDR3 proper motions ($\chi^2=3.5$ for the four proper motion difference measurements plotted). 

Figure~\ref{fig:orbithist} shows the marginalised posteriors of orbital parameters from our fit, as well as the marginalised posteriors of \citet{De_Rosa_2020_51Erib}, as retrieved from \url{whereistheplanet.com} \citep{whereistheplanet}. The angles $\Omega$, $\omega$, and $\lambda_{\rm ref}$ are bimodal, and all parameters except planet mass have non-Gaussian distributions. The planet mass posterior is well-described by a Gaussian distribution truncated at zero on the low end. The best-fit Gaussian parameters did not change significantly ($\lesssim$0.1\,\Mjup) when we tried different bin sizes or high-mass cutoffs (to exclude the less Gaussian tail). Using 0.5-\Mjup\ bins up to a maximum mass of 15\,\Mjup, we found that a Gaussian with a mean of 1.0\,\Mjup\ and standard deviation of 5.0\,\Mjup\ provided the best fit to the posterior distribution ($\chi^2=12$ for 27 degrees of freedom).

We derived marginalised, 1$\sigma$ and 2$\sigma$ confidence limits on all parameters by computing the minimum credible intervals containing 68.3\% and 95.4\% of our MCMC posterior. These are reported in Table~\ref{tbl:orbit}. We caution that these marginalised intervals should only be used to compare to future results of the same type and are not to be used for computing orbits given the significant covariances (Figure~\ref{fig:corner}).

Our orbital analysis results in parameter distributions that agree well with previous work using the same relative astrometry as we do \citep{De_Rosa_2020_51Erib} and with an independent analysis using VLT/SPHERE data \citep{Maire_2019_51Eri}. Our eccentricity posterior extends to slightly higher eccentricities than found by \citet{Maire_2019_51Eri}, but our 1$\sigma$ range overlaps with theirs.

\section{Discussion} \label{sec:models}

We find that the insignificant proper motion anomaly measured by \Hipparcos\ and \Gaia\ is in excellent agreement with independent relative astrometry measurements, rules out high planet masses, and is fully consistent with zero mass. This means that we can derive an upper mass limit for 51~Eri~b without needing to invoke the presence of other bodies in the system sufficiently massive to cause astrometric perturbations. The lack of a strong proper motion anomaly in \Gaia~EDR3 data requires a planet with a mass $<$12\,\Mjup\ with 98\% confidence (one-tailed test significance of 2.0$\sigma$). This limit is under the conservative assumption of a linear-flat mass prior, which corresponds to the higher-mass range of 2--12\,\Mjup\ having 5$\times$ the prior weight as the lower-mass range 0--2\,\Mjup.

A mass at or above 12\,\Mjup\ for 51~Eri~b has not previously been ruled out at such high significance. This is thanks to an improvement in the proper motion anomaly accuracy of 2.8$\times$ between DR2 and EDR3. \citet{De_Rosa_2020_51Erib} reported a 1$\sigma$ upper limit of 7\,\Mjup\ when using both \Hipparcos\ and \Gaia~DR2 proper motions, but they were cautious about claiming a strong upper limit. They found $\approx$2$\sigma$ disagreement between the absolute astrometry and a planet of even zero mass; their best-fit mass would have been negative without invoking another companion in the system. Our upper mass limit is thus an important milestone in showing that 51~Eri~b is most likely not a massive, cold-start, gas-giant planet. 

Furthermore, the lack of any significant proper motion anomaly is consistent with no other massive planets in the system interior to 51~Eri~b. We cannot rule out another planet on a period short compared to the $\approx$3-year \Gaia~EDR3 baseline. A planet with a 5- to 10-year period could induce a proper motion anomaly opposite to that caused by 51~Eri~b thereby reducing our measured dynamical mass.  In practice, this requires careful tuning of such a hypothetical interior planet's orbital phase, and may require an orbital resonance to be stable given 51~Eri~b's maximum likelihood $\approx$4\,AU periastron distance.

We now consider in detail the implications of cold- and hot-start models for the mass of 51~Eri~b given its age and luminosity. We can derive a posterior distribution for the mass of 51~Eri~b using substellar evolutionary models. There are several possible choices for hot-start models \citep[e.g.,][]{Burrows1997, 2008ApJ...689.1327S, 2020A&A...637A..38P}, but for the sake of self-consistency we use the hot- and cold-start versions of the \citet{2012ApJ...745..174S} models. As can be seen in their Figure~4, these models represent even more disparate initial entropy states than considered in earlier work by \citet{2007ApJ...655..541M}.

\subsection{The luminosity of 51~Eri~b}

Aside from mass and age, the key parameter in our analysis is the luminosity of 51~Eri~b. \citet{2017AJ....154...10R} derived two luminosities from fitting different model atmospheres to their Gemini/GPI spectrum and Keck/NIRC2 photometry of the planet, finding $\log(\Lbol/\Lsun) = -5.83\pmoffs{0.15}{0.12}$\,dex from atmospheres with iron-silicate clouds and $-5.93\pmoffs{0.19}{0.14}$\,dex from sulfide-salt atmospheres. Both sets of models used equilibrium chemistry, which resulted in poor fits to the mid-infrared photometry. In addition to this complication, VLT/SPHERE observations from \citet{2017A&A...603A..57S} are inconsistent with the $J$-band portion of the GPI spectrum and the photometry in the $K$ band.

We re-derived the luminosity of 51~Eri~b in a direct, photometry-based approach. We used the magnitude tables of the ATMO~2020 atmosphere models that include disequilibrium chemistry \citep{2020A&A...637A..38P}. From these we computed a bolometric correction (${\rm BC}_X \equiv M_{\rm bol} - M_X$) for each of the five standard filters for which 51~Eri~b has photometric measurements and for which ATMO~2020 reports magnitudes. These are the $J_{\rm MKO}$, $H_{\rm MKO}$, $K_{\rm MKO}$, $L^{\prime}$, and $M^{\prime}$ bands. We note that the $M_S$ filter in NIRC2 has properties $\lesssim$1\% different from the $M^{\prime}$ bandpass adopted by ATMO~2020.

We used both the ``strong'' and ``weak'' disequilibrium chemistry sets of models and restricted our analysis to model masses $<$15\,\Mjup\ and $\Teff=400$--800\,K. For each model we computed a $\chi^2$ for the measured absolute magnitudes versus the model magnitudes, and we kept the best-matching half of the models. The rms scatter of the remaining BC values ranged from 0.1\,mag in the $L^{\prime}$ band to 0.3\,mag in the $K_{\rm MKO}$ band and 0.5\,mag in the $M^{\prime}$ band. This trend follows the expected pattern that some bandpasses will vary more or less depending on underlying assumptions for chemistry and gravity.

Using the photometry from \citet{2017AJ....154...10R} we then computed luminosities from our model-derived bolometric corrections, accounting for the uncertainties in both in a Monte Carlo fashion. Adopting an additional uncertainty of 0.2\,dex (0.5\,mag), to account for systematic errors in models and/or observations, we found good agreement between the luminosities derived from each band ($\chi^2 = 5$ for five measurements). Taking the weighted average of all five luminosities we found $\log(\Lbol/\Lsun) = -5.5\pm0.2$\,dex for 51~Eri~b. This is in good agreement with the value of $-5.6$\,dex that \citet{2015Sci...350...64M} derived from assuming a low-gravity, partly-cloudy spectrum. Comparing to the 2D posteriors of \citet{2017AJ....154...10R} shown in their Figures~11 and 14, our luminosity range overlaps with their 1$\sigma$ contours and agrees best with their larger model radii, as expected for such a young substellar object. 

\subsection{Masses derived from evolutionary models}

Because we measure an upper limit on the mass, we used evolutionary models to derive posterior distributions on mass for comparison. Cold-start evolution presents a challenge for interpolating  model grids because luminosity changes very slowly even at young ages, in contrast to hot-start models with rapid initial cooling. Some cold-start models even have crisscrossing luminosity curves of different masses as a function of age that would make interpolations double-valued \citep[e.g.,][]{2008ApJ...683.1104F}. We therefore took a Monte Carlo rejection-sampling approach, where we randomly drew ages and masses to compute luminosities from all model tracks. Comparing to the observed luminosity and its uncertainty allowed for a collection of results covering the full range of modelled masses. This approach is the same as we have used previously in our work testing models with dynamical masses \citep[e.g.,][]{2017ApJS..231...15D,Brandt_2021_HR8799e}.

In our analysis we used the full grid of evolutionary models (hot- and cold-start) from \citet{2012ApJ...745..174S} that is available online (see Data Availability). These cover the most tabulated masses and ages of any cold-start models, and they also provide solar- and 3$\times$ solar-metallicity as well as a wide range of initial specific entropy values ($\Sinit$).\footnote{Following the previous literature, we discuss \Sinit\ in units of Boltzmann's constant ($k_{\rm B}$) per baryon.} \citet{Marleau+Cumming_2014} found good agreement between these models, their own, and others, so our results are not expected to depend strongly on our choice of models. 

The online \citet{2012ApJ...745..174S} model data includes a spectrum spanning 0.8--15\,\micron\ at each model grid point, with the grid spanning 8.0--13.0\,\kbb\  in \Sinit\ (steps of 0.25\,\kbb), 1--15\,\Mjup\ in mass (steps of 1\,\Mjup), and 1\,Myr to 1\,Gyr in age (ten approximately logarithmic steps per decade). We computed bolometric fluxes by numerically integrating these spectra, extrapolating to zero flux at 0\,\micron\ and assuming a Rayleigh-Jeans tail beyond 15\,\micron. We visually inspected the spectra to confirm that in almost all cases the long-wavelength end was already displaying a Rayleigh-Jeans slope. Only for the coldest models, much colder than 51~Eri~b, was this not visually apparent. The Rayleigh-Jeans flux contributes about 9\% of the bolometric flux at $\log(\Lbol/\Lsun)=-6.0$\,dex and even less at higher luminosities.

\begin{figure}
  \includegraphics[width=0.49\textwidth]{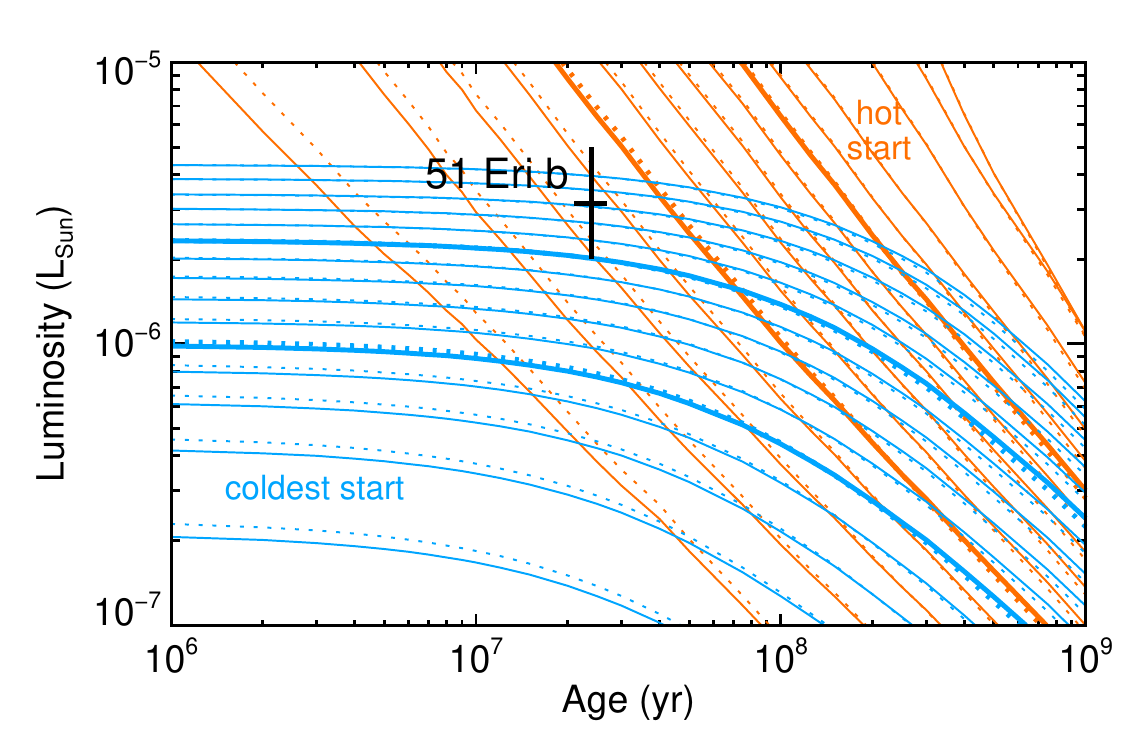}
  \caption{Evolutionary models from \citet{2012ApJ...745..174S} showing the hot-start (orange lines) and coldest-start ($\Sinit=8.0$\,\kbb; blue lines) scenarios. Every model mass is displayed, ranging from 1\,\Mjup\ to 15\,\Mjup\ in 1-\Mjup\ steps. The 5\,\Mjup\ and 10\,\Mjup\ models are highlighted by thicker plotted lines. Solid lines are solar-metallicity models, and dotted lines are 3$\times$ solar. 51~Eri~b is consistent with both scenarios but at different masses: $\lesssim$5\,\Mjup\ in a hot start and $\gtrsim$10\,\Mjup\ in the coldest start.}
  \label{fig:models}
\end{figure}

Figure~\ref{fig:models} shows the evolutionary models we used in our analysis. The 3$\times$ solar-metallicity models are negligibly different in the predicted luminosity at a given mass and age, so we used the solar-metallicity models in our analysis.

We first considered the simplest case of hot-start models, with \Sinit\ ranging from 10.75\,\kbb\ at 1\,\Mjup\ to 13.00\,\kbb\ at $\geq$10\,\Mjup. We follow the same prescription as \citet{2012ApJ...745..174S} for the maximum \Sinit\ as a function of mass, which ensures that all planets are gravitationally bound. We assumed a linear-flat prior in mass over the whole grid of models and a normally-distributed age corresponding to the BPMG \citep[$24\pm3$\,Myr;][]{2015MNRAS.454..593B}. This resulted in a mass posterior with a median and 1$\sigma$ interval of $3.8_{-0.8}^{+0.7}$\,\Mjup\ using the evolutionary tracks of \citet{2012ApJ...745..174S}. We found that other widely-used hot-start models give similar masses; for example, \citet{2008ApJ...689.1327S} models result in a planet mass of $3.1_{-0.7}^{+0.5}$\,\Mjup\ using the same analysis. 

For cold-start models, in any given model comparison we adopted a single value for \Sinit. These values ranged from the models' lowest \Sinit\ (coldest-start) case of 8.0\,\kbb\ up to an intermediate (warm-start) value of 9.5\,\kbb. The coldest-start models from \citet{2012ApJ...745..174S} reach luminosities only as high as $\log(\Lbol/\Lsun) = -5.4$\,dex at the age of the BPMG. This means that our analysis essentially had an upper limit imposed by the model grid on mass, as 32\% of the 2-D luminosity-age distribution was higher than the highest cold-start track (corresponding to a mass of 15\,\Mjup). We corrected for this in our single-tailed probability calculations by approximating that 32\% of the model-derived masses would be $>$15\,\Mjup\ if those models were available. The single-tailed test asks whether each Monte Carlo trial for the model-derived mass is higher than the mass drawn from our measured posterior distribution. So if, for example, 50\% of our trials tested positive then our corrected probability would be $(1-0.32)\times0.50 + 0.32 = 0.66$.

\begin{figure*}
 \begin{center}
  \includegraphics[width=\textwidth]{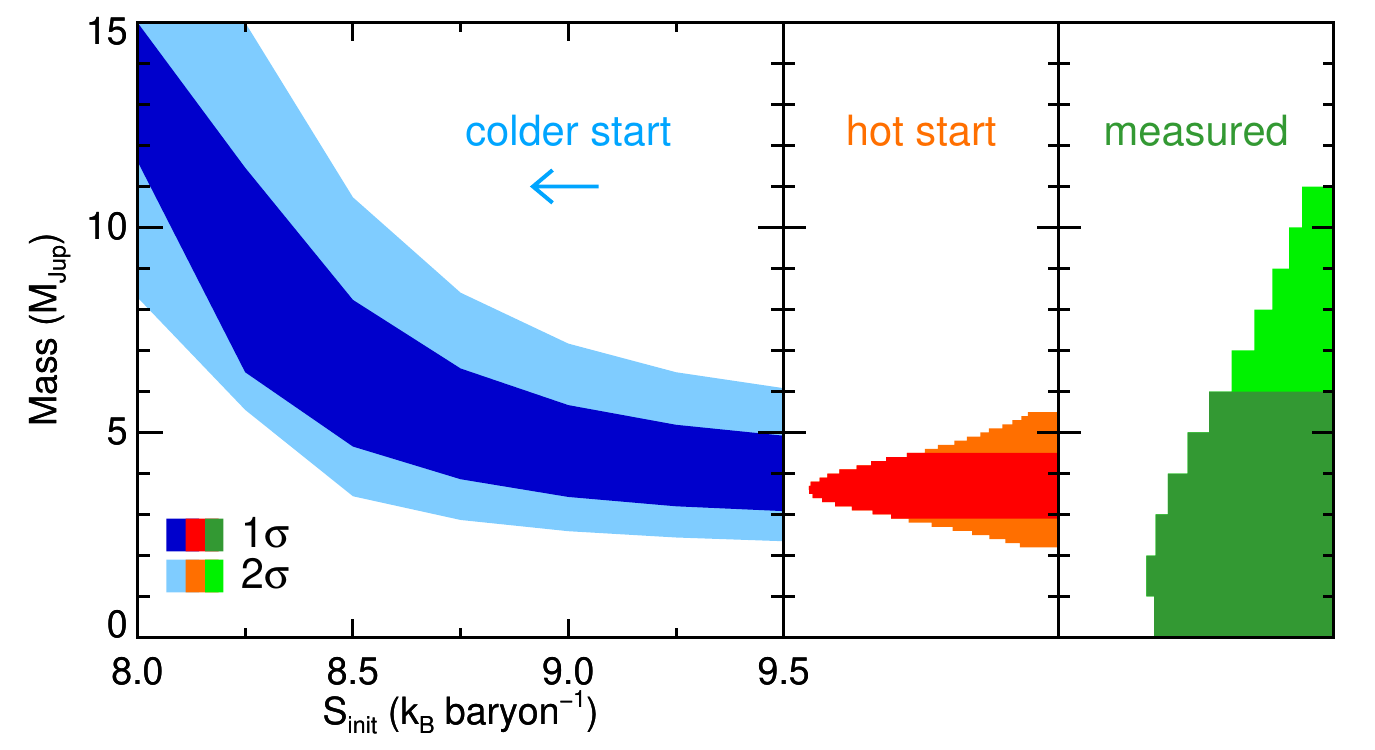} 
  \caption{Left: confidence intervals on the mass of 51~Eri~b derived from evolutionary models under different values for the cold-start initial entropy (colder starts toward the left). Middle: histogram of the hot-start evolutionary model-derived mass of 51~Eri~b that peaks around 3--4\,\Mjup. Right: histogram of our inferred mass posterior distribution using HGCA-EDR3 astrometry. In all panels, the darker shading corresponds to 1$\sigma$ confidence intervals and lighter shading corresponds to 2$\sigma$. The coldest-start scenario here is inconsistent with our measured upper limit at 97.0\% confidence.} \label{fig:mass-sinit}
 \end{center}
\end{figure*}

Figure~\ref{fig:mass-sinit} shows our inferred mass ranges for 51~Eri~b under different assumed initial entropy. As expected, for higher initial entropy the planet's luminosity can be explained by a lower mass, while conversely the lowest entropy case requires a very high mass to reproduce 51~Eri~b's luminosity. Even with a moderately warm-start entropy of $\Sinit = 9.5$\,\kbb, the mass inferred from evolutionary models is nearly the same as from the hot-start scenario.

In single-tailed tests of the model-derived mass distribution, we found that the coldest-start \citet{2012ApJ...745..174S} models ($\Sinit=8.0$\,\kbb) were inconsistent with our observed distribution at 97.0\% confidence. This drops to only 76\% confidence at the next model grid step of $\Sinit=8.25$\,\kbb.

\subsection{Implications for the origin of 51~Eri~b}

The limit of $\Sinit>8$\,\kbb\ that we find for the initial entropy of 51~Eri~b is the first such result for a potentially cold-start planet. It agrees well with such constraints derived by \citet{Marleau+Cumming_2014} for other directly imaged planets from luminosity alone. They found  $\Sinit>10.5$\,\kbb\ for $\beta$~Pic~b and $\Sinit>9.2$\,\kbb\ for the HR~8799 planets and 2MASSW~J1207334$-$393254~b. 

Our result provides further observational evidence against the cold-start scenario for giant planet formation. This scenario is commonly linked to the core-accretion model of giant planet formation \citep{1996Icar..124...62P,2000Icar..143....2B,Hubickyj_etal_2005,2007ApJ...655..541M}. However, there are growing indications that not only does the core-accretion model not necessarily imply a low initial entropy, but such cold starts may be rare in core accretion. \citet{Mordasini_2013} showed that even completely cold core accretion can lead to a high entropy and planet luminosity if the core mass is sufficiently large. \citet{2017ApJ...834..149B} concluded that cold-start core accretion may be difficult to achieve without the accretion shock staying cool (close to the temperature of the nebula), in addition to accretion rates staying low. Indeed, they suggest that the special conditions needed imply that 51~Eri~b is unlikely to be a cold-start planet, and simulations of accretion shocks by \citet{2019ApJ...881..144M} further support this idea. Therefore, core accretion is an entirely plausible formation scenario for 51~Eri~b, in spite of our result ruling out low initial entropy for the planet.

Our entropy constraint is equally consistent with a hot-start formation by gravitational instability. It is generally accepted that this mechanism is unlikely to form most giant planets \citep[e.g.,][]{2005ApJ...621L..69R}, especially those on closer orbits like 51~Eri~b with a semimajor axis of  $10.4\pmoffs{0.8}{1.1}$\,AU. 
It has also been suggested that if some gas giants do form by gravitational instability, then they are better thought of as the low-mass tail of disk-born stellar and brown dwarf companions \citep{2010ApJ...710.1375K}. 
Observations broadly support this view, as population studies find that the companion mass function rises steeply to lower masses \citep[e.g.,][]{2019ApJ...877...46W}, and these lower-mass giant planets orbit higher-metallicity stars \citep{2018ApJ...853...37S}---both hallmarks of core accretion. Our measured mass distribution for 51~Eri~b is consistent with both lower- and higher-mass planet regimes in the context of these previous works.

In order to explain the present-day location of 51~Eri~b in the gravitational instability scenario, disk migration could have brought it inward from an initially wider separation. This might also help explain its eccentricity, as only the widest planets formed by gravitational instability are expected to have as eccentric orbits as 51~Eri~b \citep{2017MNRAS.470.2517H}. However, such a formation history faces the serious problem that planets as massive as 51~Eri~b would seem to be expected to migrate outward, not inward, from their birthplace \citep{2021ApJ...918L..36D}. Thus, perhaps an alternative pathway to decrease semimajor axis and increase eccentricity for 51~Eri~b in the gravitational instability scenario would be forming on a wider orbit and being scattered inward by another planet that may or may not remain in the system at the present day. 

Indeed, even in the core accretion scenario, planet-planet scattering would probably be required to achieve the relatively high eccentricity of 51~Eri~b. \citet{Maire_2019_51Eri} discussed dynamical origins for the eccentricity of 51~Eri~b in detail and concluded that Kozai-Lidov oscillations from GJ~3305~AB, the wide stellar companion to 51 Eri, are not likely. Alternatively, secular interactions with the disk can pump up eccentricity but only to perhaps $e\sim0.2$ \citep[e.g.,][]{2001A&A...366..263P}. But even then a high planet mass ($\gtrsim$10\,\Mjup), which is disfavoured by our mass posterior, is needed to clear a big enough gap in the disk to reach the highest eccentricities. Planet-planet scattering tends to retain the more massive of the two original planets, so if the hypothetical scatterer were still in the system, it would likely be fainter than 51~Eri~b and undetectable so far.

To summarise, our upper limit on the initial entropy of 51~Eri~b rules out the coldest-start versions of the core accretion scenario. Formation by core accretion does not necessarily imply a low initial entropy, and, conversely, a high initial entropy does not necessarily imply formation by gravitational instability. Therefore, our results are consistent with a high-entropy core accretion scenario, while the $\sim$10\,AU orbit of 51~Eri~b disfavours formation by gravitational instability.

\subsection{Prospects for an improved dynamical mass}

The mass uncertainty scales linearly with the proper motion uncertainty (e.g., see Equation~7 of \citealp{2019AJ....158..140B}), which in turn scales as $t^{-1.5}$. \Gaia~DR4 is expected to include 60 months of data, compared to the 34 months in EDR3, implying a factor of 2.3 improvement with no other changes to the astrometric precision of bright stars like 51~Eri~b. Therefore, using the same approach as we have here comparing the \Gaia\ proper motion to the long-term \Hipparcos-\Gaia one is expected to provide a 2.3$\times$ more precise mass for 51~Eri~b.

To validate this expectation, we performed an identical orbital analysis using \Gaia\ proper motion errors that we reduced artificially by a factor of 2.3. Given that the input proper motion anomaly was still zero, we found an upper limit of 4.2\,\Mjup\ at 2$\sigma$, which is about a 2.5$\times$ improvement over our EDR3-based result and consistent with the expectation. If further improvements are made for bright stars, then a 1-$\sigma $mass uncertainty of $\approx$1\,\Mjup\ may be possible. In this more optimistic case, the hot-start model-predicted mass of $3.1\pmoffs{0.5}{0.7}$\,\Mjup\ would mean a 3--4$\sigma$ detection of 51~Eri~b using \Gaia~DR4.

\Gaia~DR3 will provide seven-parameter solutions that include acceleration, and \Gaia~DR4 will provide epoch astrometry. These will enable a measurement of the mass of 51~Eri~b within \Gaia\ itself without recourse to \Hipparcos. In order to determine the expected acceleration as a function of planet mass, we fitted second-order polynomials to the astrometric orbit of 51~Eri~A, as defined by our MCMC posterior, uniformly sampled in time from 2014.5 onward. For a nominal 5-year mission we find second-order terms in RA and Dec of $\ddot{\alpha^*} = 0.8\,\mu{\rm as\,yr}^{-2} \times (M_{\rm pl}/\Mjup)$ and $\ddot{\delta} = 2.1\,\mu{\rm as\,yr}^{-2} \times (M_{\rm pl}/\Mjup)$, respectively. The scatter about the fit, which we performed in log-space, was about 0.06\,dex, which reflects the uncertainty in our orbit fit. For a 10-year extended \Gaia\ mission, the second-order terms are similar in amplitude but the goodness-of-fit is worse, with scatter about the second-order polynomial of $\approx$2\,$\mu$as, as expected for Keplerian motion over a significant fraction of an orbital period not following a simple polynomial.

To estimate the accuracy of such second-order polynomial measurements in future  \Gaia\ data releases, we used the same epoch-astrometry tool {\tt htof} as in our orbit analysis. {\tt htof} can access up to 10~years of predicted scans from the \Gaia\ observation forecast tool GOST\footnote{\url{https://gaia.esac.esa.int/gost/}} and compute the precision in seven-parameter fits. To calibrate this prediction for 51~Eri, we adjusted the along-scan errors so that the fitted parameter errors for a simulation of \Gaia~EDR3 matched those in the published catalogue. It is expected that \Gaia~DR4 will be based on 5.5~years of astrometry, for which {\tt htof} predicts uncertainties in the RA and Dec second-order terms of 31\,$\mu$as\,yr$^{-2}$ and 38\,$\mu$as\,yr$^{-2}$, respectively. This would be insufficient to detect 51~Eri~b at any plausible mass. For a 10-year mission, {\tt htof} predicts an uncertainty of 6\,$\mu$as\,yr$^{-2}$ in acceleration for 51~Eri, which corresponds to a $\approx$1$\sigma$ detection of a hot-start 51~Eri~b. Overall, our analysis suggests that substantial improvements will likely be needed in the astrometric errors for bright stars like 51~Eri in \Gaia\ to achieve a significant acceleration detection without using the \Hipparcos-\Gaia\ long-term proper motion.

As for detecting 51~Eri~b by RVs, our orbital analysis predicts the semi-amplitude of the signal due to 51~Eri~b, with an upper limit of 53\,m\,s$^{-1}$ at 2$\sigma$. This will not be feasible to detect given the star's pulsation amplitude of 1476\,m\,s$^{-1}$ and $v\sin{i}$ of 80\,\kms\ \citep{2020A&A...633A..44G}. The planet $\beta$~Pic~b provides a useful point of comparison as its semimajor axis ($10.26\pm0.10$\,AU) is similar to 51~Eri~b, but its orbit is nearly edge-on and mass ($9\pm3$\,\Mjup) is much higher \citep{Brandt_2021_beta_Pic_bc}. Assuming 51~Eri~b is about 3$\times$ less massive than  $\beta$~Pic~b, its more face-on inclination implies that its RV semi-amplitude would be about 5$\times$ smaller. $\beta$~Pic~A has been observed with HARPS intensively for at least 15~years \citep{2019NatAs...3.1135L,2020AJ....160..243V} and not yet yielded a significant RV detection of $\beta$~Pic~b (unlike the closer-in planet $\beta$~Pic~c), so 51~Eri~b is expected to likewise be beyond the reach of RV detection.

\section{Conclusions}

We present an upper limit on the mass of 51~Eri~b derived from the cross-calibration of the \Hipparcos\ and \Gaia~EDR3 catalogues by \citet{Brandt_2021_HGCA_EDR3}. Our joint analysis of the new absolute astrometry and relative astrometry from the literature results in a concordant orbit fit that rules out planet masses of $\geq$10.9\,\Mjup\ at 2$\sigma$ and $\geq$12\,\Mjup\ at 98\% confidence.

We have reassessed the luminosity of 51~Eri~b using a direct, photometric approach that relies minimally on the details of atmospheric models. The luminosity of $\log(\Lbol/\Lsun) = -5.5\pm0.2$\,dex that we find is somewhat higher than reported by \citet{2017AJ....154...10R} but generally consistent with other work \citep{2015Sci...350...64M,2017A&A...603A..57S}.

We have derived a lower limit on the initial specific entropy of 51~Eri~b ($\Sinit>8.0$\,\kbb) that rules out the coldest-start planet formation scenarios. This supports 51~Eri~b forming in a similar way as other directly-imaged planets, like the $\beta$~Pic and HR~8799 planets, that are only consistent with hot- or warm-start scenarios. On its own, a moderately-high initial entropy does not necessarily favour formation via either core accretion or gravitational instability. However, 51~Eri~b's semimajor axis of $10.4\pmoffs{0.8}{1.1}$\,AU is most consistent with core accretion, perhaps with subsequent planet-planet scattering to explain its moderate eccentricity (0.35--0.66 at 2$\sigma$).

An important step toward understanding the origins of the coldest directly imaged planets like 51~Eri~b will include spectroscopic measurements to probe its composition with existing \citep[e.g., Keck/KPIC;][]{Mawet_2017_KPIC,Wang_2021_KPIC_HR8799} and upcoming ground-based facilities (e.g., Subaru/REACH; \citealp{2020SPIE11448E..78K}; and VLT/HiRISE; \citealp{2018SPIE10702E..36V}), as well as \JWST. On a similar time horizon, we expect to see some improvement in the measured mass of 51~Eri~b with future \Gaia\ data releases. Beyond \Gaia, the {\sl Nancy Grace Roman Space Telescope} has the potential to improve dynamical masses for planets like 51~Eri~b with novel techniques to enable high-precision absolute astrometry for bright stars \citep{2018AJ....155..102M,2019JATIS...5d4005W}.

%%%%%%%%%%%%%%%%%%%%%%%%%%%%%%%%%%%%%%%%%%%%%%%%%%
\section*{Acknowledgements}
We are indebted to Kaitlin Kratter and Ken Rice for very helpful discussions about planet formation theory.
We are grateful to the reviewer for a thoughtful and prompt report that improved the manuscript.
G.M.B.\ is supported by the National Science Foundation (NSF) Graduate Research Fellowship under grant \#1650114.
T.B.\ gratefully acknowledges support from NASA under grant \#80NSSC18K0439.
This work has made use of data from the European Space Agency (ESA) mission {\it Gaia} (\url{https://www.cosmos.esa.int/Gaia}), processed by the {\it Gaia} Data Processing and Analysis Consortium (DPAC, \url{https://www.cosmos.esa.int/web/Gaia/dpac/consortium}). Funding for the DPAC has been provided by national institutions, in particular the institutions participating in the {\it Gaia} Multilateral Agreement.
%%%%%%%%%%%%%%%%%%%%%%%%%%%%%%%%%%%%%%%%%%%%%%%%%%
\section*{Data Availability}

HGCA-EDR3 is available as a FITS file in the supplemental article data at \url{https://iopscience.iop.org/article/10.3847/1538-4365/abf93c}.
ATMO~2020 model data can be retrieved from \url{https://noctis.erc-atmo.eu:5001/sharing/zyU96xA6o}. The model data from \citet{2012ApJ...745..174S} can be retrieved from \url{https://www.astro.princeton.edu/~burrows/warmstart/index.html}.
Our MCMC chain and {\tt orvara} configuration file are available in the supplemental data in the online version of this article.


%%%%%%%%%%%%%%%%%%%%%%%%%%%%%%%%%%%%%%%%%%%%%%%%%%

% Don't change these lines
%\bsp	% typesetting comment
\label{lastpage}
\end{document}